\documentclass[12pt]{article}

\usepackage{amsmath,axodraw,rotating,calc,graphics,Feyngraphs,color} 

\def\theequation{\arabic{section}.\arabic{equation}}

\newcommand{\mycaption}[2][kurz]{{\begin{center} 
\parbox{15cm}{{\bf \caption[#1]{\rm {#2}}}} \end{center} }}

\newcommand{\la}{{\mathcal L}}

\newcommand{\da}{^{\dagger}}
\newcommand{\gs}{g'_5 \, \! ^2}
\newcommand{\rxi}{$R_{\xi}$}

\newcommand{\equ}[1]{{(\ref{#1})}}

\newcommand{\app}[1]{{Appendix \ref{#1}}}

\newcommand{\lr}[1]{{ \left( \, #1 \, \right) }}
\newcommand{\lreckig}[1]{{ \left[ \, #1 \, \right] }}
\newcommand{\mr}[1]{{\mathrm{#1}}}
\newcommand{\Vector}[1]{{\left( \begin{array}{c} #1 \end{array} \right) }}

\newcommand{\Psibar}{\overline{\Psi}}
\newcommand{\ov}[1]{{\overline{#1}}}

\newcommand{\sib}{\sin\beta}
\newcommand{\cob}{\cos\beta}

\newcommand{\ssb}{\sin^2\beta}
\newcommand{\csb}{\cos^2\beta}

\begin{document}

\begin{flushright}
MC-TH-2001-010\\[-0.15cm]
WUE-ITP-2001-030\\[-0.15cm] 
hep-ph/0110391\\[-0.15cm]
October 2001
\end{flushright}

\begin{center}
{\Large {\bf Minimal Higher-Dimensional Extensions of the}}\\[0.3cm]
{\Large {\bf Standard Model and Electroweak Observables}}\\[1.4cm]
{\large  Alexander M\"uck$^{\, a}$, Apostolos Pilaftsis$^{\, a,b}$ 
and Reinhold R\"uckl$^{\,a}$}\\[0.4cm]
$^a${\em Institut f\"ur Theoretische Physik und Astrophysik,
         Universit\"at W\"urzburg,\\ Am Hubland, 97074 W\"urzburg, 
         Germany}\\[0.2cm]
$^b${\em Department of Physics and Astronomy, University of Manchester,\\
         Manchester M13 9PL, United Kingdom}
\end{center}
\vskip1.cm \centerline{\bf ABSTRACT} 
We   consider minimal 5-dimensional   extensions of the Standard Model
compactified   on an $S^1/Z_2$  orbifold, in   which the SU(2)$_L$ and
U(1)$_Y$ gauge fields and Higgs bosons may or may not all propagate in
the fifth  dimension while the  observable matter is always assumed to
be confined to a  4-dimensional subspace.  We pay particular attention
to    consistently  quantize  the   higher-dimensional  models  in the
generalized $R_\xi$ gauge and derive analytic expressions for the mass
spectrum  of the resulting  Kaluza-Klein states and their couplings to
matter.   Based  on recent data  from  electroweak precision tests, we
improve previous limits  obtained in the 5-dimensional Standard  Model
with a common compactification radius and extend our analysis to other
possible 5-dimensional Standard-Model constructions.  We find that the
usually   derived lower bound    of   $\sim 4$~TeV  on  an   universal
compactification scale may be considerably relaxed  to $\sim 3$~TeV in
a minimal scenario,  in which the  SU(2)$_L$ gauge  boson is the  only
field that feels the presence of the fifth dimension.

\newpage

\setcounter{equation}{0}
\section{Introduction}

\indent

String theory provides the   only known theoretical   framework within
which gravity  can  be quantized and  so  undeniably plays   a central
r\^ole in our endeavours of unifying all fundamental forces of nature.
A  consistent  quantum-mechanical   formulation  of  a   string theory
requires the  existence of additional dimensions  beyond the four ones
we experience in our  every-day life.  These new  dimensions, however,
must  be  sufficiently  compact so  as  to  escape detection.  In  the
original   string-theoretic  considerations~\cite{review}, the inverse
length $1/R$ of the extra compact dimensions and the string mass $M_s$
turned out to be closely tied to the 4-dimensional Planck mass $M_{\rm
  P} = 1.9\times 10^{16}$~TeV, with  all involved mass scales being of
the    same    order.      More  recent     studies,   however,   have
shown~\cite{IA,JL,EW,ADS,DDG}  that  there could  still be conceivable
scenarios  of  stringy nature  where $1/R$ and  $M_s$   may be lowered
independently of $M_{\rm  P}$ by several  or many orders of magnitude. 
In particular, Ref.~\cite{ADS} considers  the radical possibility that
$M_s$ is of order TeV and represents the only fundamental scale in the
universe at which unification of all forces of nature occurs.  In this
model, the  compactification radius  related to the higher-dimensional
gravitational  interactions lies  in  the sub-millimeter range,  i.e.\ 
$1/R  \stackrel{<}{{}_\sim} 10^{-3}$~eV, so Cavendish-type experiments
may potentially  test the model by  observing deviations from Newton's
law~\cite{ADS,KS}  at such small  distances.  The  model also offers a
wealth    of    phenomenological   implications   for      high-energy
colliders~\cite{GRW}.

The above  low string-scale framework  could be embedded  within e.g.\
type I string theories~\cite{EW}, where the Standard Model (SM) may be
described    as   an    intersection   of    higher-dimensional   $Dp$
branes~\cite{ADS,DDG,AB}.   As  such  intersections may  naturally  be
higher dimensional, in addition to gravitons the SM gauge fields could
also  propagate independently  within  a higher-dimensional  subspace,
where the size of the new  extra dimensions is of order TeV$^{-1}$ for
phenomenological  reasons~\cite{NY,WJM,CCDG,DPQ2,RW,DPQ1}.  Since such
low  string-scale constructions  may effectively  result  in different
higher-dimensional  extensions  of  the SM~\cite{AB,AKT},  the  actual
limits  on the  compactification  radius are,  to  some extent,  model
dependent.   Nevertheless,  in  the  existing literature  the  derived
phenomenological limits  were obtained by  assuming that the  SM gauge
fields propagate  all freely in a common  higher-dimensional space, in
which  the compactification  radius  is universal  for  all the  extra
dimensions.

In this paper we  wish to lift the   above restriction and  extend the
analysis  to models which  minimally  depart from  the assumption of a
universal higher-dimensional scenario.  Specifically, we will consider
5-dimensional extensions of   the  SM  compactified on  an   $S^1/Z_2$
orbifold, where the SU(2)$_L$  and U(1)$_Y$ gauge  bosons may not both
live in  the  same higher-dimensional space,   the so-called bulk. For
example, one  could imagine that the  SU(2)$_L$ gauge field propagates
in the   bulk  whilst the U(1)$_Y$   gauge  boson is  confined  to our
observable  4-dimensional  subspace and vice  versa.   This observable
4-dimensional subspace is often termed  3-brane or simply brane and is
localized at one  of the two fixed  points of the $S^1/Z_2$  orbifold,
the boundary.  In the aforementioned higher-dimensional scenarios, all
SM fermions and the Higgs  boson  should necessarily be brane  fields,
such that an explicit breaking  of the 4-dimensional gauge symmetry of
the original (classical) Lagrangian is avoided.

Another issue of particular  interest to us is  related to our ability
of consistently quantizing the  higher-dimensional models under  study
in the so-called $R_\xi$  gauge.  In particular, it  can be shown that
higher-dimensional  gauge-fixing conditions  can  always be found that
reduce to the  usual $R_\xi$ gauge after  the compact dimensions  have
been integrated   out.    Such  a   quantization   procedure   can  be
successfully   applied to both  Abelian and  non-Abelian theories that
include Higgs   bosons living in  the bulk  and/or on the  brane.  The
$R_\xi$   gauge  has the   attractive  theoretical  feature  that  the
unphysical  sector  decouples from   the theory in  the  limit  of the
gauge-fixing parameter $\xi \to \infty$, thereby allowing for explicit
checks  of  the gauge  independence  of physical observables,  such as
S-matrix elements.

After compactification of the extra dimensions, we obtain an effective
4-dimensional theory which is  usually described by infinite towers of
massive Kaluza--Klein (KK) states.  In the 5-dimensional extensions of
the SM  under consideration, such infinite  towers generically consist
of  KK excitations  of the  $W$-boson, the  $Z$-boson and  the photon.
Since the mass  of the first excited KK state is  typically set by the
inverse of  the compactification radius  $R$, one expects that  the KK
effect  on  high-precision electroweak  observables  will become  more
significant for  higher values of $R$.   Thus, if all  SM gauge bosons
live  in  the bulk,  compatibility  of  this  model with  the  present
electroweak  data gives  rise to  a lower  bound~\cite{CCDG}  of $\sim
4$~TeV on $1/R$ at the 2$\sigma$ level.

On the other hand, the possibilities that the SU(2)$_L$ gauge boson is
a brane  field with the  U(1)$_Y$ gauge boson  living in the  bulk and
vice  versa are  phenomenologically  even more  challenging.  In  such
cases,  we find  that the  lower limit  on the  compactification scale
$1/R$ can become significantly weaker, i.e.\ $1/R\stackrel{>}{{}_\sim}
3$~TeV.  This new result emerges  partially from the fact that some of
the  most  constraining  high-precision  electroweak  observables  are
getting differently affected  by the presence of the  KK states within
these mixed  brane-bulk scenaria. For example, the  muon lifetime does
not  directly receive  contributions from  KK excitations  if  the $W$
boson lives on the brane, but only indirectly when the analytic result
is expressed  in terms  of the  $Z$-boson mass in  the context  of our
adopted  renormalization scheme.   Most interestingly,  unlike  in the
frequently investigated  model with all  SM gauge fields in  the bulk,
other competitive  observables, such  as $A^b_{\rm FB}$  and $A^e_{\rm
LR}$~\cite{PDG},   do  now   possess   additional  distinct   analytic
dependences  on the  compactification scale  $1/R$ within  these novel
brane-bulk  models. As  a consequence,  the results  of  the performed
global-fit analysis become substantially different for these scenaria.

The  paper  is  organized as  follows:  in  Section  2 we  consider  a
5-dimensional Abelian model compactified  on an $S^1/Z_2$ orbifold, in
which the gauge field propagates  in the bulk.  The model is quantized
by  prescribing the  proper higher-dimensional  gauge-fixing condition
which  leads to  the usual  class of  $R_\xi$ gauges  after  the extra
dimension has  been integrated  out.  The same  gauge-fixing procedure
may successfully  be implemented for  Abelian models augmented  by one
Higgs boson which could either be a bulk or a brane field, or even for
more general  models with two Higgs  bosons where the  one Higgs boson
can live on the brane and the  other one in the bulk.  In Section~2 we
also present  analytic expressions for  the masses of the  physical KK
gauge  bosons  and  for  their  mixings with  the  corresponding  weak
eigenstates.  In  Section~3 we extend our gauge-fixing  procedure to a
higher-dimensional non-Abelian theory  and discuss the basic structure
of  the gauge sector  after compactification.   In Section~4  we study
5-dimensional  extensions  of  the  SM,  in which  the  SU(2)$_L$  and
U(1)$_Y$  gauge  fields and  Higgs  bosons may  or  may  not all  feel
simultaneously  the  presence  of  the  compact  dimension  while  the
fermionic matter  is always assumed to  be confined on  the brane.  In
fact,  we distinguish  three cases:  (i) both  SU(2)$_L$  and U(1)$_Y$
gauge bosons are bulk fields, (ii)  only the U(1)$_Y$ gauge boson is a
bulk field  while the SU(2)$_L$ one  is a brane field,  and (iii) only
the SU(2)$_L$ gauge  boson resides in the bulk  while the U(1)$_Y$ one
is restricted to the brane.   Technical details of our study have been
relegated  to the  Appendices  A and  B.   In Section~5  we perform  a
fully-fledged global-fit analysis  to the aforementioned 5-dimensional
extensions  of  the  SM,   based  on  recent  data  on  high-precision
electroweak observables.  Section~6 summarizes our conclusions.

\setcounter{equation}{0}
\section{5-Dimensional Abelian Model}\label{5DQED}

To describe  as well  as motivate our  higher-dimensional gauge-fixing
quantization  procedure, it is  very instructive  to consider  first a
simple  Abelian  5-dimensional model,  such  as 5-dimensional  Quantum
Electrodynamics  (5D-QED)   where  the  extra   spatial  dimension  is
compactified  on an  $S^1/Z_2$ orbifold.   Then, we  shall  extend our
quantization procedure to more general Abelian models with bulk and/or
brane Higgs fields.

As a starting point, let us consider the 5D-QED Lagrangian given by
\begin{equation}
\label{freelagrangian}
\la (x, y) \, = \, - \frac{1}{4} F_{M N} (x, y) F^{M N} (x, y) \:
+\: \la_{\mr{GF}}(x,y)\: +\: \la_{\rm FP}(x,y)\, ,
\end{equation}
where
\begin{equation}
\label{fieldstrength}
F_{M N} (x, y) \, = \, \partial_M A_N (x, y) - \partial_N A_M (x, y)
\end{equation}
denotes    the    5-dimensional    field    strength    tensor,    and
$\la_{\mr{GF}}(x,y)$ and $\la_{\rm  FP}(x,y)$ are the gauge-fixing and
the   induced  Faddeev--Popov  ghost   terms,   respectively.   In   a
5-dimensional Abelian  model, one may neglect the Faddeev--Popov ghost
term $\la_{\rm FP}$  induced by $\la_{\rm GF}$, as  the Abelian ghosts
are  non-interacting   and  hence   they  cannot  occur   in  S-matrix
elements. We shall return to  this point in Section~3, when discussing
quantization of higher-dimensional non-Abelian theories.

Throughout the paper, Lorentz indices in 5 dimensions are denoted with
capital Roman  letters, e.g.~$M,N  = 0,1,2,3,5$, while  the respective
indices  pertaining to  the ordinary  4 dimensions  are  symbolized by
Greek  letters, e.g.~$\mu,\nu  =  0,1,2,3$.  Furthermore,  we use  the
abbreviations  $x  =  (x^0,\vec{x})$  and  $y =  x^5$  to  denote  the
coordinates  of the  usual $1+3$-dimensional  Minkowski space  and the
coordinate of the fifth compact dimension, respectively.

In a 5-dimensional theory, the gauge-boson field $A_M$ transforms as a
vector  under  the Lorentz  group  SO(1,4).   In  the absence  of  the
gauge-fixing  and ghost  terms $\la_{\rm  FP}$ and  $\la_{\rm  GF}$ in
(\ref{freelagrangian}),  the 5D-QED  Lagrangian is  invariant  under a
U(1) gauge transformation:
\begin{equation}
\label{gaugetrf}
A_M(x,y) \to A_M(x,y) + \partial_M \Theta(x,y) \, .
\end{equation}
Being consistent  with the  above property of  gauge symmetry,  we can
compactify  the  theory  on  an  $S^1/Z_2$  orbifold,  such  that  the
following equalities are satisfied:
\begin{equation}
\label{fieldconstraints}
\begin{split}
A_M(x,y) \, & = \, A_M(x,y + 2 \pi R)\,, \\ 
A_{\mu}(x,y) \, & = \, A_{\mu}(x, - y)\,, \\ 
A_5(x,y) \, & = \, - A_5(x, - y)\,,\\ 
\Theta(x,y) \, & = \, \Theta(x,y + 2 \pi R)\,,\\ 
\Theta(x,y) \, & = \, \Theta(x, - y)\, .
\end{split}
\end{equation}
As we will see below, the fact  that $A_\mu (x,y)$ is taken to be even
under  $Z_2$ results  in  the  embedding of  conventional  QED with  a
massless photon into our 5D-QED.  Notice that all other constraints on
the  field  $A_5(x,y)$  and  the  gauge parameter  $\Theta  (x,y)$  in
(\ref{fieldconstraints})  follow  automatically if  the  theory is  to
remain gauge invariant after compactification.

Given the periodicity and  reflection properties of $A_M$ and $\Theta$
under $y$ in~(\ref{fieldconstraints}),  we can expand these quantities
in a Fourier series as follows:
\begin{equation}
\label{fourierseries}
\begin{split}
A^{\mu}(x, y) \, & = \, \frac{1}{\sqrt{2 \pi R}} \, A^{\mu}_{(0)}(x)
                     \, + \, \sum_{n=1}^{\infty} \, \frac{1}{\sqrt{\pi
                     R}} \, A^{\mu}_{(n)}(x) \, \cos \lr{ \frac{n
                     y}{R}} \, , \\ 
A^5(x, y) \, & = \, \sum_{n=1}^{\infty} \, \frac{1}{\sqrt{\pi R}} \,
                     A^5_{(n)}(x) \, \sin \lr{ \frac{n y}{R}} \, , \\
\Theta (x, y) \, & = \, \frac{1}{\sqrt{2 \pi R}}
                     \, \Theta_{(0)}(x) \, + \, \sum_{n=1}^{\infty} \,
                     \frac{1}{\sqrt{\pi R}} \, \Theta_{(n)}(x) \cos
                     \Bigl( \frac{n y}{R} \Bigr) \, .
\end{split}
\end{equation}
The Fourier  coefficients $A_{(n)}^{\mu}  (x)$, also called  KK modes,
turn out to be the mass eigenstates in 5D-QED.  However, this is not a
generic feature of higher-dimensional models, namely the Fourier modes
cannot always  be identified  with the KK  mass eigenstates.  Below we
will encounter examples,  in which the Fourier modes  will mix to form
the KK mass eigenstates.

{}From \equ{gaugetrf} and \equ{fourierseries},  one can now derive the
corresponding gauge transformations for the KK modes~\cite{DDG}
\begin{equation}
\label{gaugetrfinmodes}
\begin{split}
A_{(n) \mu}(x) \, & \to \, A_{(n) \mu}(x) + \partial_{\mu}
\Theta_{(n)}(x) \, , \\ 
A_{(n) 5}(x) \, & \to \, A_{(n) 5}(x) -
\frac{n}{R} \Theta_{(n)}(x) \, .
\end{split}
\end{equation}
Integrating out  the $y$ dimension yields  the effective 4-dimensional
Lagrangian
\begin{eqnarray}
\label{labeforegf}
\la(x) &=& -\, \frac{1}{4} F_{(0) \mu \nu} \, F_{(0)}^{\mu \nu} + \sum_{n
   = 1}^{\infty} \, \Big[\, -\frac{1}{4} \, F_{(n) \mu \nu} \,
   F_{(n)}^{\mu \nu}\nonumber\\
&&+\, \frac{1}{2} \lr{ \frac{n}{R} \, A_{(n) \mu} +
   \partial_{\mu} A_{(n) 5} }\lr{ \frac{n}{R} \, A^\mu_{(n)} +
   \partial^\mu A_{(n)5} }\, \Big]\: +\:  \la_{\mr{GF}}(x) \, ,
\end{eqnarray}
where $\la_{\mr{GF}}(x) = \int_0^{2  \pi R} dy \, \la_{\mr{GF}}(x,y)$.
Note  that  the  invariance  of  $\la(x)$  under  the  transformations
\equ{gaugetrfinmodes}   becomes  manifest  in   the  absence   of  the
gauge-fixing term $\la_{\mr{GF}}(x,y)$.

In  addition to  the  usual  QED terms  involving  the massless  field
$A^{\mu}_{(0)}$,  the  other  terms  in  the  effective  4-dimensional
Lagrangian  \equ{labeforegf} describe two  infinite towers  of massive
vector   excitations   $A^{\mu}_{(n)}$   and   (pseudo)-scalar   modes
$A^{5}_{(n)}$ that  mix with  each other, for  $n \ge 1$.   The scalar
modes $A^{5}_{(n)}$ play the r\^ole of the would-be Goldstone modes in
a  non-linear realization  of an  Abelian  Higgs model,  in which  the
corresponding Higgs fields are taken to be infinitely massive.

As  in usual  Higgs  models,  one may  seek  for a  higher-dimensional
generalization  of 't-Hooft's  gauge-fixing condition,  for  which the
mixing  terms  bilinear   in  $A^{\mu}_{(n)}$  and  $A^{5}_{(n)}$  are
eliminated        from        the       effective        4-dimensional
Lagrangian~\equ{labeforegf}.  For instance, the covariant gauge-fixing
term~\cite{DDG}
\begin{equation}
\label{lorentzgauge}
\la_{\mr{GF}}(x,y) \ =\ - \, \frac{1}{2 \xi} \, \lr{\partial_M \, A^M}^2
\end{equation}
does not  lead to  a complete cancellation  of the  bilinear operators
$A^{\mu}_{(n)}\partial_\mu   A^{5}_{(n)}$   in~\equ{labeforegf},  with
exception the Feynman  gauge $\xi = 1$. Taking,  however, advantage of
the  fact  that  orbifold  compactification generally  breaks  SO(1,4)
invariance~\cite{GGH}, one  can abandon the  requirement of covariance
of the gauge fixing condition with respect to the extra dimension.  In
this  context,  we are  free  to  choose  the following  non-covariant
generalized $R_\xi$  gauge:\footnote[1]{For a related  suggestion made
recently, see~\cite{GNN}.}
\begin{equation}
\label{gengaugefixterm}
\la_{\mr{GF}}(x, y)\ =\ -\, \frac{1}{2 \xi} (\partial^{\mu} A_{\mu}
\: - \: \xi \, \partial_5 A_5)^2 \, .
\end{equation}
Nevertheless, the gauge-fixing  term in \equ{gengaugefixterm} is still
invariant under ordinary  4-dimensional Lorentz transformations.  Upon
integration over the extra dimension,  it is not difficult to see that
all mixing terms  involving $A^{\mu}_{(n)}\partial_\mu A^{5}_{(n)}$ in
\equ{labeforegf} drop  out up to  irrelevant total derivatives.   As a
consequence,  the  propagators  for  the  fields  $A^{\mu}_{(n)}$  and
$A^{5}_{(n)}$ take  on their usual  forms that describe  massive gauge
fields and  their respective would-be Goldstone bosons  of an ordinary
4-dimensional Abelian-Higgs model in the $R_\xi$ gauge:
\begin{center}
\begin{picture}(460,090)(60,40)
\SetWidth{1.}
\thicklines
\Scalar{160, 50}{220, 50}
\put(183,58){$(n)$} 
\put(275,48){$ = 
\frac{i}{k^2 - \xi \left( \frac{n}{R} \right)^2}$}
\Photon(160,100)(220,100){3}{5} 
\put(148,98){$\mu$} 
\put(183,108){$(n)$} 
\put(228,98){$\nu$}
\put(275,98){$= 
\frac{i}{k^2 - \left( \frac{n}{R} \right)^2} \, 
\lreckig{ - g^{\mu \nu} \, + \, \, 
\frac{ \left( 1 - \xi \right) k^{\mu} \, k^{\nu}}{ 
k^2 - \xi \left( \frac{n}{R} \right)^2} \, }$} 
\stepcounter{equation}
\put(490,98){(\arabic{section}.\arabic{equation})}
\stepcounter{equation}
\put(490,48){(\arabic{section}.\arabic{equation})}
\end{picture}
\end{center}
Therefore,  we  shall  often  refer  to the  $A^{5}_{(n)}$  fields  as
Goldstone modes,  even though  these KK modes  do not  directly result
from a mechanism of spontaneous symmetry breaking in the usual sense.

Having defined the appropriate  $R_\xi$ gauge through the gauge-fixing
term in~\equ{gengaugefixterm}, we can  recover the usual unitary gauge
in the limit  $\xi \to \infty$.  This limit is  also equivalent to the
gauge-fixing   condition   $A_5(x,y)   =   0$   or   equivalently   to
$A^{5}_{(n)}(x) = 0$, where all  unphysical KK scalar modes are absent
from the theory~\cite{PS}.   Thus, for the case at  hand, we have seen
how  starting  from  a non-covariant  higher-dimensional  gauge-fixing
condition, we can arrive  at the known covariant 4-dimensional $R_\xi$
gauge  after  compactification.   As  we  will see  below,  the  above
quantization   procedure   can   be   extended   to   more   elaborate
higher-dimensional  models that  may include  brane and/or  bulk Higgs
fields.

\subsection{Abelian Model with a Bulk Higgs Boson}
\label{bulkhiggs}

Here,  we shall  discuss an  extension of  the Abelian  model outlined
above by adding a bulk Higgs  scalar. The 5D Lagrangian of this theory
reads
\begin{equation}
\label{lagrdens5Dabelianhiggsmodel}
\la (x,y) \ = \ - \, \frac{1}{4} \, F^{M N}\, F_{M N}\: +\:
               (D_M\Phi)^*\, (D^M \Phi ) \: - \: V(\Phi )
               \: + \: \la_{\mr{GF}}(x,y) \, ,
\end{equation}
where $D_M  = \partial_M \,  + \,  i \, e_5  \, A_M$ is  the covariant
derivative,  $e_5$  denotes  the  5-dimensional  gauge  coupling,  and
$\Phi(x,y)$ is the 5-dimensional complex scalar field
\begin{equation}
\label{higgsfieldbulkmodel}
\Phi (x,y)\ = \ 
\frac{1}{\sqrt{2}} \, \Big(\, h(x,y) \, + \, i \, \chi(x,y)\, \Big)
\end{equation}
that transforms under a U(1) gauge transformation as
\begin{equation}
\label{gaugetrfhiggsinthebulokabelian}
\Phi (x,y) \quad \to \quad 
\exp\Big( \!- i\, e_5 \, \Theta(x,y)\Big)\: \Phi (x,y) \, .
\end{equation}
In   \equ{lagrdens5Dabelianhiggsmodel},    the   5-dimensional   Higgs
potential is given by
\begin{equation}
\label{Higgspotentialinfivedim}
V(\Phi) \, = \, \mu_5^2 \, | \Phi |^2 \, + \, \lambda_5 \, | \Phi |^4\, , 
\end{equation}
with $\lambda_5 > 0$.

After imposing the $S^1/Z_2$ compactification conditions $\Phi (x,y) =
\Phi (x,y + 2\pi R)$ and $\Phi (x,y) = \Phi (x,-y)$ on $\Phi(x,y)$, we
can perform  a Fourier decomposition of  the scalar fields  $h (x, y)$
and $\chi (x,y)$ in terms of cosines
\begin{equation}
\label{hchiexpansion}
\begin{split}
h(x,y)\, & = \, \frac{1}{\sqrt{2 \pi R}} \, h_{(0)}(x) \, 
          + \,\sum_{n=1}^{\infty} \, \frac{1}{\sqrt{\pi R}}\, 
                        h_{(n)}(x) \, \cos \lr{\frac{n y}{R}} \, , \\
\chi(x,y)\, & = \, \frac{1}{\sqrt{2 \pi R}} \, \chi_{(0)}(x) \, + \,  
           \sum_{n=1}^{\infty}\, \frac{1}{\sqrt{\pi R}}\, 
          \chi_{(n)}(x) \, \cos \lr{\frac{n y}{R} } \, .
\end{split}
\end{equation}
As we will see below, our choice  of an even $Z_2$ parity for the bulk
Higgs scalar $\Phi$ ensures that  the lowest lying KK modes describe a
conventional  4-dimensional Abelian Higgs  model.  Instead,  if $\Phi$
were odd under $Z_2$, this  would not allow Yukawa interactions of the
Higgs  scalars  with fermions  localized  on  a  brane $y=0$  and  the
generation  of fermion  masses through  the Higgs  mechanism  would be
impossible in this case.

Let us  now turn our  attention to the  effective Higgs sector  of our
Abelian model.  The effective 4-dimensional Lagrangian associated with
the Higgs fields may conveniently be given by
\begin{equation}
\label{kintermsforhiggsinabmodel}
\la_{\mr{Higgs}} (x) \ = \ \frac{1}{2} \, \sum_{n=0}^{\infty}\Big[\,
(\partial_{\mu} h_{(n)})\,(\partial^{\mu} h_{(n)})\: -\:
\frac{n^2}{R^2}\, h_{(n)}^2\: - \: 
\mu^2 \, h_{(n)}^2\: + \: \lr{h \leftrightarrow \chi} \Big] \ +\ \dots\,,
\end{equation}
where $\mu^2  = \mu^2_5$ and the ellipses  denote quartic interactions
which involve  the Higgs fields  $h_{(n)}$ and $\chi_{(n)}$  and which
all   depend   on   $\lambda   =   \lambda_5/(2\pi  R)   >   0$.    In
\equ{kintermsforhiggsinabmodel},   the  mass  terms   proportional  to
$n^2/R^2$ arise from compactifying the $y$-dimension.  As in the usual
4-dimensional  case,  for  $\mu^2  <   0$,  the  zero  KK  Higgs  mode
$\Phi_{(0)}   =  (h_{(0)}   +  i   \chi_{(0)})/\sqrt{2}$   acquires  a
non-vanishing vacuum expectation value (VEV)
\begin{equation}
\label{vevdef}
\langle \Phi_{(0)} \rangle \, = \, \frac{1}{\sqrt{2}} \, 
\langle h_{(0)} \rangle \, = \, \frac{v}{\sqrt{2}},
\end{equation}
which  breaks the U(1)  symmetry.  Moreover, it   can be shown that as
long as  the phenomenologically relevant condition $v  < 1/R$  is met,
$h_{(0)}$ will be the only mode to receive  a non-zero VEV, i.e.\ $v =
\sqrt{|\mu|^2/\lambda}$.

After spontaneous symmetry breaking, the effective kinetic Lagrangian
of the theory for the $n$-KK mode may be cast into the form:
\begin{eqnarray}
\label{searchforgoldstone}
\la^{(n)}_{\rm kin}(x) &=& \, -\, \frac{1}{4}\, F_{(n)}^{\mu \nu} \,
                          F_{(n) \mu \nu} \ + \ 
\frac{1}{2}\, \Big(\, \frac{n^2}{R^2}\: +\: e^2v^2\Big) A_{(n) \mu}
                                              A_{(n)}^{\mu}\: + \: 
\frac{1}{2}\, (\partial_{\mu} A_{(n) 5})\,
                                 (\partial^{\mu} A_{(n)5})\nonumber\\ 
&&+\, \frac{1}{2}\, (\partial_{\mu} \chi_{(n)})(\partial^{\mu} \chi_{(n)})
\ - \ \frac{1}{2} \Big(\, \frac{n}{R}\, \chi_{(n)}
\: -\: ev\, A_{(n) 5}\Big)^2\nonumber\\
&& +\, A_{(n)}^{\mu} \, \partial_{\mu}\, \Big(\, \frac{n}{R}\, A_{(n) 5}
\: +\: ev \, \chi_{(n)}\Big)\ +\ \dots
\end{eqnarray}
where $e = e_5 / \sqrt{2 \pi R}$ and the dots indicate the omission of
bilinear terms  involving $h_{(n)}$.  {}From \equ{searchforgoldstone},
it is evident that the mass spectrum of the zero KK modes is identical
to that of a conventional Abelian  Higgs model, i.e.\ $m_{A (0)} = ev$
and $m_{h (0)} = \sqrt{2 \lambda}  v$. This is so, because $A_{(0) 5}$
is absent and we are  left with the standard 4-dimensional terms only.
To determine  the complete mass spectrum  for the higher  KK modes, we
first introduce  the (pseudo)-scalar  KK modes $G_{(n)}$  and $a{(n)}$
through the orthogonal linear transformations:
\begin{equation}
\label{physunphysfieldshiggsinbulk}
\begin{split}
G_{(n)} \, & = \, \Big(\, \frac{n^2}{R^2}\: +\: e^2v^2\,
\Big)^{-1/2}\, \lr{ \frac{n}{R} \, A_{(n) 5} \: 
+\: ev\,\chi_{(n)} } \, , \\
a_{(n)} \, & = \, \Big(\, \frac{n^2}{R^2}\: +\: e^2v^2\,
\Big)^{-1/2}\, 
\lr{\, ev\, A_{(n) 5} \: - \: \frac{n}{R}\, \chi_{(n)}} \, .
\end{split}
\end{equation}
Then,   with    the   aid     of    \equ{physunphysfieldshiggsinbulk},
$\la^{(n)}_{\rm kin}$ in  \equ{searchforgoldstone} can be rewritten in
the more compact form
\begin{equation}
  \label{Lkineff}
\begin{split}
\la^{(n)}_{\rm kin} (x)\, = & \, 
-\, \frac{1}{4} \, F_{(n)}^{\mu \nu} \, F_{(n) \mu \nu} \, 
+ \, \frac{1}{2}\, \big( m_{A (n)} \, A_{(n) \mu} \, + \, 
                              \partial_{\mu} \, G_{(n)}\big)\,
\big( m_{A (n)} \, A^\mu_{(n)} \, + \, 
                              \partial^{\mu} \, G_{(n)}\big)  \\
   & \, \, + \, \frac{1}{2}\, (\partial_\mu a_{(n)}) \,
(\partial^\mu a_{(n)})\: - \: 
\frac{1}{2} m^2_{a (n)} a_{(n)}^2 \ +\ \dots\, ,
\end{split}
\end{equation}
with  $m^2_{A (n)} = m^2_{a (n)}  =  (n^2/R^2) + e^2v^2$.  {}From this
last  expression   for $\la^{(n)}_{\rm  kin}$,    we readily  see that
$G_{(n)}$ plays the  role of  a  Goldstone mode  in  an Abelian  Higgs
model, while the pseudoscalar field $a_{(n)}$  describes a physical KK
excitation  degenerate in mass with the  KK  gauge mode~$A_{(n) \mu}$.
In particular, since the zero  KK modes of the  fields are expected to
be much  lighter than their  first KK  excitations, i.e.\ $ev\ll 1/R$,
the  masses   of  all  higher   $n$-KK  gauge  and   Higgs  modes  are
approximately $m_{(n)}  = n/R$ and the  Goldstone  modes $G_{(n)}$ may
almost be  identified with $A_{(n)5}$, i.e.\ $G_{(n)}\approx A_{(n)5}$
as in 5D-QED.

{}From the above discussion, it becomes now clear that the appropriate
gauge-fixing  Lagrangian  in  \equ{lagrdens5Dabelianhiggsmodel}  for a
5-dimensional generalized \rxi-gauge should be
\begin{equation}
\label{gaugefixingfunctionAbelianmodel}
\la_{\mr{GF}} (x,y) \, = \, - \, \frac{1}{2 \xi} \, \bigg[\,\partial_{\mu}
                         A^{\mu} \, -  \, \xi \, \bigg(\,\partial_5 A_5
           \: + \: e_5 \frac{v}{\sqrt{2 \pi R}} \, \chi\,\bigg)\,\bigg]^2 \, .
\end{equation}
Taking \equ{gaugefixingfunctionAbelianmodel} into account, we   arrive
at the total effective kinetic Lagrangian
\begin{equation}
\label{gaugefixedlaabbulk}
\begin{split}
\la^{(n)}_\mr{kin} (x)\ =\ & 
- \frac{1}{4} \, F_{(n)}^{\mu \nu} \, F_{(n) \mu \nu} 
        \: + \: \frac{1}{2} \, m_{A (n)}^2 \, A_{(n) \mu}\,A^\mu_{(n)}
\: - \: \frac{1}{2 \xi} (\partial_{\mu} \, A_{(n)}^{\mu})^2 \\
& + \, \frac{1}{2} 
(\partial_{\mu} G_{(n)})\,(\partial^{\mu} G_{(n)})\:
- \: \frac{\xi}{2} \, m_{A (n)}^2 \, G_{(n)}^2\\
    & + \, \frac{1}{2} (\partial_{\mu} a_{(n)})\,(\partial^{\mu} a_{(n)})
\: - \: \frac{1}{2} \, m_{a (n)}^2 \, a_{(n)}^2 \\
&+\, \frac{1}{2} (\partial_{\mu} h_{(n)})\,(\partial^{\mu} h_{(n)})\: 
- \: \frac{1}{2}\, m_{h (n)}^2 \, h_{(n)}^2 \, .
\end{split}
\end{equation}
In the above,  $m_{h (n)} = \sqrt{(n^2/R^2)  + 2 \lambda v^2}$ are the
KK Higgs masses and $m_{A (n)}$  and $m_{a (n)}$  are the KK masses of
$A_{(n)}$ and $a_{(n)}$  given  after \equ{Lkineff}.  Observe  finally
that   the  limit    $\xi   \to  \infty$   in~\equ{gaugefixedlaabbulk}
consistently corresponds to the unitary gauge.

\subsection{Abelian Model with a Brane Higgs Boson}
\label{branehiggs}

A  qualitatively different way of  implementing the  Higgs sector in a
higher-dimensional Abelian model is to localize the Higgs field at the
$y=0$ boundary of the $S^1/Z_2$ orbifold. The 5-dimensional Lagrangian
of this theory reads
\begin{equation}
\label{lagrdens5Dabelianhiggsmodelbrane}
\la (x,y)\ =\ - \, \frac{1}{4} \, F^{M N}\, F_{M N}\: +\: 
\delta (y) \, \lreckig{ (D_{\mu} \Phi)^* \,(D^\mu \Phi)  \: 
                  - \: V(\Phi )}\: +\:  \la_{\mr{GF}}(x,y) \, .
\end{equation}
Here,  the  covariant derivative  $D_{\mu} =   \partial_{\mu}  + i e_5
A_{\mu} (x,y)$ and the Higgs potential  $V(\Phi) = \mu^2  | \Phi |^2 +
\lambda |  \Phi |^4$ have  their familiar 4-dimensional forms, and the
$\delta$-function $\delta (y)$ confines the Higgs  sector on the brane
$y=0$.  Under a gauge  transformation, the brane Higgs field $\Phi(x)$
transforms as
\begin{equation}
\label{gaugetrfabelianhiggsmodelhiggsonbrane}
\Phi (x) \quad \to \quad 
\exp\Big( \!- i\, e_5 \, \Theta(x,0)\Big)\: \Phi (x) \, .
\end{equation}
Under \equ{gaugetrfabelianhiggsmodelhiggsonbrane}   and     the  local
transformation~\equ{gaugetrf} of    the gauge  field   $A_M(x,y)$, the
theory exhibits U(1)  invariance. Notice  that  the bulk  scalar field
$A_5 (x,y)$ vanishes  on the  brane $y =  0$  as a  result of its  odd
$Z_2$-parity.

After  compactification and integration   over the $y$-dimension,  the
effective Lagrangian of the model under discussion will  be the sum of
two terms: the effective Lagrangian~\equ{labeforegf} of 5D-QED and the
square         bracket                     $[\ldots]_{y=0}$         in
\equ{lagrdens5Dabelianhiggsmodelbrane}.   Obviously, $\Phi=(h+i\chi) /
\sqrt{2}$ being a brane field does not possess KK excitations and, for
$\mu^2 <  0$  (with $\lambda  >  0$), acquires  a VEV:  $\langle  \Phi
\rangle  = \langle  h  \rangle /   \sqrt{2} =  v  / \sqrt{2}$.   After
spontaneous symmetry  breaking, masses  are  generated for  all the KK
gauge modes $A^{\mu}_{(n)}$. However, unlike in the Abelian model with
a bulk Higgs    boson  discussed in  Section~2.1,  the   corresponding
gauge-boson mass matrix here is no longer diagonal and has the form
\begin{equation}
\label{massmatrixvectorbosonshiggsonbrane}
M_{A}^2 \, = \, 
\left(
\begin{array}{cccc}
m^2 & \sqrt{2} \, m^2 & \sqrt{2} \, m^2 & \cdots \\
\sqrt{2} \, m^2 & 2 \, m^2 \, + \, (1/R)^2 & 2 \, m^2 & \cdots \\
\sqrt{2} \, m^2 & 2 \, m^2 & 2 \, m^2 \, + \, (2/R)^2 & \cdots \\
\vdots & \vdots & \vdots & \ddots
\end{array}
\right) \quad ,
\end{equation}
where  $m^2  = e^2  v^2$ denotes  the   mass generated  by  the  Higgs
mechanism.  The eigenvalues of $M_A^2$ follow from
\begin{equation}
\label{characpolynphyshiggsonbrane}
\det \lr{M_{A}^2 \, - \, \lambda \, \mr{I}} \, = \,
\lr{\prod_{n=1}^{\infty} \lr{ n^2/R^2 - \lambda }} \, \lr{ m^2 \, - \,
\lambda \, - \, 2 \, \lambda \, m^2 \sum^{\infty}_{n=1} \,
\frac{1}{\lr{n / R}^2 \, - \, \lambda}} \, = \, 0 \, .
\end{equation}
Since  $\lambda \, = \,  (n/R)^2$ is not a solution   as can be easily
seen, the mass eigenvalues  $m_{A(n)}$ are given by  the zeros of  the
second  big bracket   in  \equ{characpolynphyshiggsonbrane}.   This is
equivalent to solving the transcendental equation
\begin{equation}
\label{transcendentalequformasses}
\sqrt{\lambda} \, = \, m_{(n)} \, = \, \pi \, m^2 \, R \, \cot \lr{\pi
\, m_{(n)} \, R} \, ,
\end{equation}
with $m_{A(n)}  =   m_{(n)}$.   The  respective  KK  mass  eigenstates
$\hat{A}^{\mu}_{(n)}$ are given by
\begin{equation}
\label{masseigenstatesexplicit}
\hat{A}^{\mu}_{(n)} \ = \ \bigg( 1 \, +\, \pi^2 \, m^2 \, R^2 +
\frac{m^2_{(n)}}{m^2}\bigg)^{- 1/2} \, \sum^{\infty}_{j=0} \,
\frac{2 \, m_{(n)} \, m}{m^2_{(n)} \, - \, (j / R)^2} \,
\bigg(\frac{1}{\sqrt{2}}\bigg)^{\delta_{j,0}} \, A^{\mu}_{(j)} \, .
\end{equation}

To    find the  appropriate     form     of the  gauge-fixing     term
$\la_{\mr{GF}}(x,y)$   in~\equ{lagrdens5Dabelianhiggsmodelbrane},   we
follow \equ{gaugefixingfunctionAbelianmodel},  but restrict the scalar
field $\chi$ to the brane $y = 0$, viz. 
\begin{equation}
\label{gaugefixingfunctionabelianhiggsonbrane}
\la_{\mr{GF}} (x,y) \, = \, - \, \frac{1}{2 \xi} \, 
        \Big[\, \partial_{\mu} A^{\mu} \ - \ 
\xi \, \big( \partial_5 \, A_5 \: + \: 
e_5 v \, \chi \, \delta(y)\big)\,\Big]^2 \, .
\end{equation}
Then,      the   effective   4-dimensional   gauge-fixing   Lagrangian
$\la_{\mr{GF}}(x)$ is given by
\begin{equation}
\label{gftermsinlaHiggsonbrane}
\begin{split}
\la_{\mr{GF}} (x) \, = \, & - \, \frac{1}{2 \xi} \, \big(\partial_{\mu}
                              A^{\mu}_{(0)}\big)^2 \, - \, \frac{1}{2 \xi}
                              \, \sum_{n=1}^{\infty} \,
                              \lr{\partial_{\mu} A^{\mu}_{(n)} \, - \,
                              \xi \, \frac{n}{R} \, A_{(n) 5}}^2 \\ \,
& + \, e v \, \chi \, \big(\partial_{\mu}
                              A^{\mu}_{(0)}\big) \, + \, \sqrt{2} \, e v
                              \, \chi \, \sum^{\infty}_{n=1} \,
                              \big(\partial_{\mu} A^{\mu}_{(n)}\big)\\ 
& - \, \xi \, \sqrt{2} \, e v \, \chi \,
                              \sum^{\infty}_{n=1} \, \frac{n}{R}\,
                              A_{(n) 5} \, - \, \frac{\xi}{2} \, e_5^2
                              v^2 \chi^2 \delta(0) \, .
\end{split}
\end{equation}
On the $S^1/Z_2$ orbifold, the $\delta$-function may be represented by
\begin{equation}
\label{deltafunctionrepres}
\delta(y) \, = \, \frac{1}{2 \pi R} \, + \, \sum^{\infty}_{n=1} \,
\frac{1}{\pi R} \, \cos \lr{\frac{n y}{R}}\, ,
\end{equation}
which implies
\begin{equation}
\label{deltazerorepres}
\delta(0) \, = \, \frac{1}{2 \pi R} \, + \, \sum^{\infty}_{n=1} \,
\frac{1}{\pi R} \, .
\end{equation}
It is  interesting to  verify  whether our  5-dimensional gauge-fixing
term in \equ{gaugefixingfunctionabelianhiggsonbrane} does consistently
lead to the generalized $R_\xi$ gauge after integration over the extra
dimension. In doing so, we apply the $R_\xi$-gauge-fixing prescription
individually to    each KK gauge mode    in the  effective Lagrangian,
instead   of using \equ{gaugefixingfunctionabelianhiggsonbrane}. It is
then not difficult to obtain
\begin{equation}
\la^{(n)}_{\mr{GF}} (x) \ = \ 
- \, \frac{1}{2 \xi} \, \Big[\, \partial_{\mu} A_{(n)}^{\mu} \: - \: 
\xi \, \lr{ \frac{n}{R}\, A_{(n) 5} \,
  + \, \sqrt{2}^{\lr{1 - \delta_{n,0}}} e v \, \chi }\, \Big]^2\, .
\end{equation}
This    analytic      result   coincides  with      the     one stated
in~\equ{gftermsinlaHiggsonbrane}, provided $e_5 = \sqrt{2 \pi R} \, e$
and  \equ{deltazerorepres}  are  used.  As  is also   expected  from a
generalized $R_\xi$     gauge, all mixing  terms   of  the gauge modes
$A_{(n)}^{\mu}$ with $A_{(n)  5}$ and  $\chi$  disappear up  to  total
derivatives.    Hence,   the eigenvalues $m_{(n)}$   as   derived from
\equ{transcendentalequformasses} do represent the physical masses.

The unphysical mass spectrum of  the Goldstone modes may be determined
by diagonalizing the    following $\xi$-dependent mass matrix   of the
fields $\chi$ and~$A_{(n) 5}$:
\begin{equation}
\la^{\xi}_{\mr{mass}}(x)\ =\ -\, \frac{\xi}{2} \left( \chi , \, A_{(1)
5} , \, A_{(2) 5} , \, \ldots \right) M_{\xi}^2 \left(
\begin{array}{c}
\chi \\ 
A_{(1) 5} \\
A_{(2) 5} \\
\vdots
\end{array}
\right)\,,
\end{equation}
with
\begin{equation}
\label{massmatrixxidependenthiggsonbrane}
M_{\xi}^2 \, = \, 
\left(
\begin{array}{cccc}
e^2 v^2 \, \lr{1 + \sum^{\infty}_{n=1} \, 2} & \sqrt{2} \, \lr{1/R} \,
e v & \sqrt{2} \, \lr{2/R} \, e v & \cdots \\ \sqrt{2} \, \lr{1/R} \,
e v & \lr{1/R}^2 & 0 & \cdots \\ \sqrt{2} \, \lr{2/R} \, e v & 0 &
\lr{2/R}^2 & \cdots \\ \vdots & \vdots & \vdots & \ddots
\end{array}
\right) \, .
\end{equation}
It can be  shown that the characteristic  polynomial of $M_{\xi}^2$ is
formally   identical     to  the     one   of     $M^2_A$   given   in
\equ{characpolynphyshiggsonbrane}:
\begin{equation}
\label{characpolynunphyshiggsonbrane}
\det \lr{M_{\xi}^2 \, - \, \lambda \, \mr{I}}\ = \ 
\det \lr{M_A^2 \, - \, \lambda \, \mr{I}} \, .
\end{equation}
Consequently,  the  mass   eigenvalues  of $M_{\xi}^2$   are given  by
$m_{(n)}$ in~\equ{transcendentalequformasses}.   Thus, as  is expected
from an $R_\xi$  gauge, we find an  one-to-one correspondence  of each
physical vector mode of mass $m_{(n)}$ to an unphysical Goldstone mode
with gauge-dependent    mass $\sqrt{\xi}  \,  m_{(n)}$.  Moreover, the
Goldstone mass eigenstates are given by
\begin{equation}
\label{wouldbeGoldstonemasseigenstatesonehiggsonbrane}
\hat{G}_{(n)}\ =\ \bigg(\, 1 \, + \, \pi^2 \, m^2 \, R^2 +
\frac{m^2_{(n)}}{m^2} \bigg)^{-1/2}  
\, \bigg(\, \sqrt{2} \, \chi \ + \ 
\sum^{\infty}_{j=1} \, \frac{2 \, (j/R) \, m}{m^2_{(n)} \, - \,
(j / R)^2} \, A_{(j) 5}\,\bigg) \, .
\end{equation}
In the  unitary gauge $\xi \to\infty$, the  fields $\hat{G}_{(n)}$, or
equivalently the  fields $A_{(n) 5}$  and $\chi$, are absent  from the
theory. Therefore, as opposed to  the bulk-Higgs model of Section 2.1,
the  present  brane-Higgs model  does  not  predict  other KK  massive
scalars apart from the physical Higgs boson $h$.

\subsection{Abelian 2-Higgs Model}
\label{twoHiggs}

It  is now interesting  to  consider a  model with  two  complex Higgs
fields: one Higgs field $\Phi_1(x,y)$ propagating in  the bulk and the
other field $\Phi_2    (x)$  localized  on a  brane   at  $y=0$.   The
5-dimensional Lagrangian of this Abelian 2-Higgs model is given by
\begin{eqnarray}
\label{lageneralabelianhiggsmodel}
\la (x,y) &=& - \, \frac{1}{4} \, F^{M N} \, F_{M N}\: + \: 
(D_M \, \Phi_1)^*\,(D^M \, \Phi_1) \: + \: 
\delta (y) \, (D_{\mu} \, \Phi_2)^*\, (D^{\mu} \, \Phi_2)\nonumber\\
&&-\, V(\Phi_1,\Phi_2) \, + \, \la_{\mr{GF}}(x,y) \, ,
\end{eqnarray}
where $V$  is   the most general  Higgs  potential  allowed   by gauge
invariance
\begin{eqnarray}
\label{generalhiggspotential}
V(\Phi_1,\Phi_2) \!\!\!&=&\!\! \mu_1^2 \, ( \Phi_1\da \Phi_1 ) \, + \, 
\lambda_1 \, ( \Phi_1\da \Phi_1 )^2 \,
                            + \, \delta(y) \, \Big[\, 
\frac{1}{2}\,\mu_2^2 \, ( \Phi_2\da \Phi_2 ) \, 
+ \, m_{12}^2 \, ( \Phi_1\da \Phi_2 )\, + \, 
\frac{1}{2}\,\lambda_2 \, ( \Phi_2\da \Phi_2 )^2 \nonumber\\ 
&+&\!\!\!  
\frac{1}{2}\,\lambda_3 \, ( \Phi_1\da \Phi_1 ) ( \Phi_2\da \Phi_2 ) \, 
   + \, \frac{1}{2}\,\lambda_4 \, ( \Phi_1\da \Phi_2 ) ( \Phi_2\da \Phi_1 )\, 
   + \, \lambda_5\, ( \Phi_1\da \Phi_2 )^2\, 
   + \, \lambda_6\, ( \Phi_1\da \Phi_1 ) ( \Phi_1\da \Phi_2 )\nonumber\\
&+&\!\!\! \lambda_7 \, ( \Phi_2\da \Phi_2 ) ( \Phi_1\da \Phi_2 ) \,
   + \, {\rm h.c.}\, \Big] \, .
\end{eqnarray}
Note that all terms involving  the brane field $\Phi_2$ are multiplied
by  a  $\delta$-function.  Here,  we  shall  restrict  ourselves to  a
CP-conserving   Higgs  sector,   i.e.\   the  parameters   $m^2_{12}$,
$\lambda_5$,        $\lambda_6$        and       $\lambda_7$        in
\equ{generalhiggspotential}  are real.   Furthermore,  we assume  that
both complex scalar fields acquire  real VEV's.  Thus, we may linearly
expand $\Phi_1$ and $\Phi_2$ around their VEV's as follows:
\begin{eqnarray}
  \label{VEV1}        \Phi_1(x,y)        &=&        \frac{1}{\sqrt{2}}
\lr{\frac{v_{1}}{\sqrt{2 \pi R}} + h_1(x,y)  + \, i \, \chi_1(x,y)} \,,\\  
\label{phitwodef}  
\Phi_2(x)  &=& \frac{1}{\sqrt{2}}  \,  \Big(\,
v_{2} + h_2(x) + \, i \, \chi_2(x)\,\Big) \, .
\end{eqnarray}
Adopting the   commonly  used notation in    2-Higgs models, we define
$v_{1} = v \cos \beta$ and $v_{2} = v \sin \beta$, i.e.\ $\tan \beta =
v_2/v_1$.

In this 5-dimensional Abelian 2-Higgs model, the effective mass matrix
$M^2_{A}$ of the  Fourier modes $A^{\mu}_{(n)}$ is  given by  a sum of
two matrices:
\begin{equation}
  \label{M2A}
M^2_{A}  \  =\    M^2_{\mr{brane}}\:  +\:      M^2_{\mr{bulk}}\, .
\end{equation}
The first matrix $M^2_{\mr{brane}}$, which includes the KK masses, may
be  obtained  by      \equ{massmatrixvectorbosonshiggsonbrane}   after
replacing  $m^2=e^2v^2$ with $m^2=e^2  v^2 \sin^2  \beta$.  The second
matrix $M^2_{\mr{bulk}}$ is proportional  to unity, $M^2_{\mr{bulk}} =
e^2 v^2  \cos^2 \beta\, {\rm  I}$. Because of the particular structure
of $M^2_A$ in  this model, the mass eigenvalues  of the KK gauge modes
are given by
\begin{equation}
m^2_{A(n)}\ =\ m^2_{(n)}\: +\: \Delta m_{(n)}^2\,,
\end{equation}
where $\Delta m_{(n)}^2 = e^2 v^2 \cos^2  \beta$ and $m_{(n)}$ are the
roots           of               the       transcendental     equation
\equ{transcendentalequformasses}. The corresponding  mass  eigenstates
$\hat{A}^{\mu}_{(n)}$     may    in    turn     be   determined     by
\equ{masseigenstatesexplicit},   after  $m_{(n)}^2$ has  been replaced
with   $m_{A (n)}^2 -  \Delta  m_{(n)}^2$.

Following a similar $R_\xi$-gauge-fixing prescription as above, we may
eliminate  the mixing terms  between   $A^{\mu}_{(n)}$ and the  fields
$A_{(n) 5}$, $\chi_{1 (n)}$ and $\chi_2$ by choosing
\begin{equation}
\label{gaugefixingfunctiongenabelianmodel}
\la_{\mr{GF}} (x,y) \ = \ - \, \frac{1}{2 \xi} \,
\bigg[\,\partial_{\mu} A^{\mu} \, - \, \xi \, \bigg(\,\partial_5 \, A_5
\, + \, e_5 \frac{v}{\sqrt{2 \pi R}} \cos \beta \, \chi_1 \, + \, e_5
v \sin \beta \, \chi_2 \, \delta(y)\bigg)\,\bigg]^2 \, .
\end{equation}
In \app{Goldstonemodes},  we show  that the resulting  Goldstone modes
$\hat{G}_{(n)}$  in  this  model  have masses  $\sqrt{\xi}  m_{A(n)}$.
These  Goldstone  modes  may  be  expressed  in  terms  of  the  other
pseudoscalar  fields  $A_{(n)  5}$,  $\chi_{1 (n)}$  and  $\chi_2$  as
follows:
\begin{equation}
\label{wouldbegoldstonemodesgeneralabelianhiggs}
\hat{G}_{(n)} \, = \, E^{(n)}_{\chi_2} \, \chi_2 \, + \,
                                    \sum_{j=0}^{\infty} \, \lr{
                                    E^{(n)}_{\chi_{1 (j)}} \, \chi_{1
                                    (j)} \, + \, E^{(n)}_{A_{(j) 5}}
                                    \, A_{(j) 5} }\, ,
\end{equation}
with
\begin{equation}
\begin{split}
E^{(n)}_{\chi_2} = \frac{1}{N} \, , \quad & \quad E^{(n)}_{A_{(j) 5}}
= - \frac{1}{N} \frac{\sqrt{2} e v \sib (j/R)}{(j/R)^2 + e^2 v^2 \csb
- m^2_{A (n)}} \, , \\ 
E^{(n)}_{\chi_{1 (0)}} = - \frac{1}{N} \frac{e^2
v^2 \sib \cob}{e^2 v^2 \csb - m^2_{A (n)}} \, , \quad & \quad
E^{(n)}_{\chi_{1 (j)}} = - \frac{1}{N} \frac{\sqrt{2} e^2 v^2 \sib
\cob}{(j/R)^2 + e^2 v^2 \csb - m^2_{A (n)}}
\end{split}
\end{equation}
and 
\begin{equation}
N^2 \, = \, \frac{1}{2} \, \frac{m^2_{A (n)}}{m^2_{A (n)} - 
     e^2 v^2 \csb} \, \bigg(\, 1 + \pi^2 e^2 v^2 \ssb \, R^2 +
     \frac{m^2_{A (n)} - e^2 v^2 \csb}{e^2 v^2 \ssb}\,\bigg) \, .
\end{equation}
The masses of the lowest-lying KK Higgs scalars strongly depend on the
details of the  Higgs  potential,  whereas  the masses of  the  higher
$n$-KK Higgs modes are approximately $n/R$.

We conclude  this section by remarking that  even for the most general
Abelian case, an appropriate higher-dimensional gauge-fixing condition
analogous  to~\equ{gaugefixingfunctiongenabelianmodel}  can  always be
found that leads after compactification  to the usual $R_\xi$ gauge as
known from ordinary 4-dimensional theories. In the following, we shall
see how the above gauge-fixing  quantization procedure can be extended
to non-Abelian models as well.

\setcounter{equation}{0}
\section{Higher-Dimensional Non-Abelian Theory}
\label{Non-Abelian}

In this section, we shall consider a pure  non-Abelian theory, such as
5-dimensional Quantum Chromodynamics (5D-QCD), without interactions to
matter.  The 5D-QCD Lagrangian takes on the simple general form
\begin{equation}
\label{nala}
\la(x, y)\ =\  -\, \frac{1}{4}\, F^a_{M N} F^{a M N} \: + \:
\la_{\mr{GF}} \: + \: \la_{\mr{FP}} \, ,
\end{equation}
where
\begin{equation}
F^a_{M N} \, = \, \partial_{M}A^a_{N} \, - \, \partial_{N}A^a_{M} \, +
\, g_5 f^{abc} A^b_{M} A^c_{N} 
\end{equation}
and $f^{abc}$ are the structure  constants of the gauge group SU($N$),
with   $N=3$  for  5D-QCD.    In~\equ{nala},  the   gauge-fixing  term
$\la_{\mr{GF}}$    and    the    induced   Faddeev--Popov   Lagrangian
$\la_{\mr{FP}}$ will be determined later in this section.

As we did for  the Abelian case, we compactify  each of the  $N$ gauge
fields $A^a_{M}(x, y)$ separately on $S^1/Z_2$ through the constraints
\equ{fieldconstraints}.   Moreover,      under   a    SU($N$)    gauge
transformation, $A^a_{M}(x,y)$ transforms as
\begin{equation}
\label{nagaugetrfin5D}
A^a_{M}(x, y) \quad \to \quad A^a_{M}(x, y) \, + \, \partial_{M}
\Theta^a(x, y) - g_5 f^{abc} \Theta^b(x, y) A^c_{M}(x, y) \, .
\end{equation}
After  a  Fourier expansion  of  $A^a_{\mu}(x,y)$, $A^a_{5}(x,y)$  and
$\Theta^a(x,y)$ according  to \equ{fourierseries}, one  finds that the
local      SU($N$)     transformation~\equ{nagaugetrfin5D}     amounts
to~\cite{DDG}
\begin{eqnarray}
  \label{finalgaugetrfpropafterrescalinginnonabth}
A^a_{(0) \mu} & \to & \, A^a_{(0) \mu} \, + \, \partial_{\mu}
\Theta^a_{(0)} \, - \, \frac{1}{2} \, \frac{g_5}{\sqrt{2 \pi R}} \,
f^{abc} \, \sum_{m=0}^{\infty}\, 2^{1 - \delta_{m,0}}\, \Theta^b_{(m)}\, 
( 1 + \delta_{m,0})\, A^c_{(m) \mu} \, ,\nonumber\\ 
A^a_{(n) \mu} & \to & \, A^a_{(n) \mu} + \partial_{\mu}
\Theta^a_{(n)} \nonumber\\
&& -\,\frac{1}{2} \frac{g_5}{\sqrt{2 \pi R}} f^{abc}
\sum_{m=0}^{\infty}\, \sqrt{2}^{1 - \delta_{m,0}}\,\Theta^b_{(m)} 
 \Big[\,  \sqrt{2}^{-\delta_{m,n}}\,( 1 + \delta_{m,n})\,
 A^c_{(|m - n|) \mu}\: +\: A^c_{(m + n) \mu}\,\Big]\, ,\nonumber\\ 
A^a_{(n) 5} & \to & \, A^a_{(n) 5} \, - \, \frac{n}{R}
\Theta^a_{(n)} \nonumber\\
&& - \, \frac{1}{2} \, \frac{g_5}{\sqrt{2 \pi R}}\,
f^{abc} \, \sum_{m=0}^{\infty} \, \sqrt{2}^{1 - \delta_{m,0}}\,
\Theta^b_{(m)}  \Big( \, \text{sgn} (n - m) \, A^c_{(|m - n|) 5} \: 
+ \: A^c_{(m + n) 5} \, \Big) \, ,
\end{eqnarray}
where $n \ge 1$.  As opposed to the Abelian case, the new feature here
is that  the  KK modes can  now  mix with  each  other  under a  gauge
transformation.  As a  result of this mixing,  any attempt to truncate
the theory at a given KK mode  $n=n_{\rm trunc}$ will explicitly break
gauge    invariance.\footnote[2]{To overcome  this difficulty,  recent
papers~\cite{Arkani01,Hill01} suggested to  match the truncated theory
with  a manifestly gauge-invariant non-Abelian chiral-type Lagrangian.
Although the two theories agree   well for $n\ll n_{\rm trunc}$,  they
have a significantly different  mass spectrum close to the  truncation
energy scale, i.e.~for KK modes $n \approx n_{\rm trunc}$.}

\begin{figure}[fp]
\begin{center}
\begin{picture}(462,250)(0,20)
\SetWidth{1.}

\thicklines \Scalar{130,20}{160,50} \Scalar{130,80}{160,50}
\Photon(160,50)(210,50){3}{5} \put(0,48){vertex with 2 scalars:}
\put(180,58){$(k)$} \put(145,20){$( \, l \, )$} \put(145,74){$(m)$}
\put(218,48){$c, \, \mu$} \put(245,48){$$} \put(260,48){$ \, g \,
\left( \frac{1}{\sqrt{2}} \right)^{(\delta_{k,0} + 1)} \,
\tilde{\delta}_{l,k,m} \, f^{abc} \lr{ p - k
}^{\mu}$} \put(120,10){$b$} \put(120,84){$a$} \Vect{131,32}{1, 1}{10}
\put(126, 38){$p$} \Vect{131,70}{1, -1}{10} \put(126, 56){$k$}
\Vect{195,43}{-1, 0}{15} \put(185, 33){$q$}

\Photon(130,120)(160,150){3}{5} \Photon(130,180)(160,150){3}{5}
\Scalar{160,150}{210,150} \put(0,148){vertex with 1 scalar:}
\put(180,158){$(k)$} \put(145,120){$( \, l \, )$} \put(145,174){$(m)$}
\put(218,148){$c$} \put(245,148){$$} \put(260,163){$ - \, i \, g \,
f^{abc} \, g^{\mu \nu} \left[ \lr{\frac{m}{R}}
\lr{\frac{1}{\sqrt{2}}}^{(\delta_{l,0} + 1)} \, \tilde{\delta}_{k,l,m}
\right. $} \put(309,133){$ \left. \, - \, \lr{\frac{l}{R}}
\lr{\frac{1}{\sqrt{2}}}^{(\delta_{m,0} + 1)} \, \tilde{\delta}_{k,m,l}
\right]$} \put(110,110){$b, \, \nu $} \put(110,184){$a, \, \mu $}
\Vect{131,132}{1, 1}{10} \put(126, 138){$p$} \Vect{131,170}{1, -1}{10}
\put(126, 156){$k$} \Vect{195,143}{-1, 0}{15} \put(185, 133){$q$}

\Photon(130,220)(160,250){3}{5} \Photon(130,280)(160,250){3}{5}
\Photon(160,250)(210,250){3}{5} \put(0,248){3-boson vertex:}
\put(180,258){$(k)$} \put(145,220){$( \, l \, )$} \put(145,274){$(m)$}
\put(218,248){$c, \, \rho$} \put(245,248){$$} \put(260,273){$\, g \,
\left( \frac{1}{\sqrt{2}} \right)^{(\delta_{k,0} + \delta_{l,0} +
\delta_{m,0} + 1)} \, \delta_{k,l,m} \,$} \put(290,248){$ \, f^{abc}
\left[ g^{\mu \nu} \left( k - p \right)^{\rho} \right. $}
\put(320,228){$ + \, g^{\nu \rho} \left( p - q \right)^{\mu} $}
\put(320,208){$ + \, \left. g^{\rho \mu} \left( q - k \right)^{\nu}
\right]$} \put(110,210){$b, \, \nu $} \put(110,284){$a, \, \mu $}
\Vect{131,232}{1, 1}{10} \put(126, 238){$p$} \Vect{131,270}{1, -1}{10}
\put(126, 256){$k$} \Vect{195,243}{-1, 0}{15} \put(185, 233){$q$}

\end{picture}
\end{center}
\mycaption{\label{feynrules3gauge} 
Feynman rules for the triple gauge boson coupling. $\delta_{k,l,m}$ and
$\tilde{\delta}_{l,k,m}$ are defined in \equ{shorthandnotation}.}
\end{figure}

\begin{figure}[fp]
\begin{center}
\begin{picture}(462,150)(0,20)
\SetWidth{1.}

\thicklines \Photon(130,20)(160,50){3}{5}
\Photon(130,80)(160,50){3}{5} \Scalar{160,50}{190,20}
\Scalar{160,50}{190,80} \put(0,48){vertex with 2 scalars:}
\put(180,58){$( \, l \, )$} \put(158,20){$(m)$} \put(145,74){$(k)$}
\put(122,36){$(n)$} \put(245,48){$$} \put(260,61){$\, i \, g^2 \,
\left( \frac{1}{\sqrt{2}} \right)^{(\delta_{k,0} + \delta_{n,0})} \,
\tilde{\delta}_{k,n,l,m} \,$} \put(270,36){$ 2 \, g^{\mu \nu} \, 
\lreckig{ f^{ace} f^{bde} \, + \, f^{ade} f^{bce} }$} \put(110,10){$b,
\, \nu$} \put(110,84){$a, \, \mu$} \put(193,10){$d$} \put(193,84){$c$}

\Photon(130,120)(160,150){3}{5} \Photon(130,180)(160,150){3}{5}
\Photon(160,150)(190,120){3}{5} \Photon(160,150)(190,180){3}{5}
\put(0,148){4-boson vertex:} \put(180,158){$( \, l \, )$}
\put(158,120){$(m)$} \put(145,174){$(k)$} \put(122,136){$(n)$}
\put(245,148){$$} \put(260,173){$ \, -i g^2 \delta_{k,l,m,n} \,
\left( \frac{1}{\sqrt{2}} \right)^{(\delta_{k,0} + \delta_{l,0} +
\delta_{m,0} + \delta_{n,0} + 2)} \,$} \put(270,148){$ \left[ f^{ace} 
f^{bde} \, \lr{ g^{\mu \nu} g^{\rho \sigma} \, - \, g^{\mu \sigma} 
g^{\nu \rho} } \right. $} \put(264,128){$ + \, f^{abe} f^{cde} \, 
\lr{ g^{\mu \rho} g^{\nu \sigma} \, - \, g^{\mu \sigma} g^{\nu \rho} } $}
\put(264, 108){$ \left. + \, f^{ade} f^{bce} \, \lr{ g^{\mu \nu} g^{\rho
\sigma} \, - \, g^{\mu \rho} g^{\nu \sigma} } \right] $}
\put(110,110){$b, \, \nu$} \put(110,184){$a, \, \mu$}
\put(193,110){$d, \, \sigma$} \put(193,184){$c, \, \rho$}

\end{picture}
\end{center}
\mycaption{\label{feynrules4gauge} 
Feynman rules for the quartic gauge boson coupling. $\delta_{k,l,m,n}$ and
$\tilde{\delta}_{k,n,l,m}$ are defined in \equ{shorthandnotation2}.}
\end{figure}

It is  straightforward to generalize the  gauge-fixing term  of 5D-QED
given in~\equ{gengaugefixterm} to  the  5D-QCD case.  The gauge-fixing
term in 5D-QCD is given by
\begin{equation}
\label{rxigaugefixtermnonab}
{\cal L}_{\rm GF}(x, y)\ =\ 
- \frac{1}{2 \xi}\, \big( F^a (A^a)\big)^2\,,
\end{equation}
with
\begin{equation}
  \label{FQCD}
F^a (A^a)\ =\ \partial^{\mu} A^a_{\mu}\: -\: \xi\, \partial_5 A^a_5 \, .
\end{equation}
In this generalized    $R_\xi$ gauge, all mixing terms   $A^a_{(n)\mu}
\partial^\mu A^a_{(n)5}$ disappear,   so  the  Fourier modes represent 
mass eigenstates. As in the Abelian case, the latter is  spoiled by  a
Higgs mechanism involving brane interactions.

In   non-Abelian theories, the   $R_\xi$  gauge induces an interacting
ghost sector, which is described by the Faddeev--Popov Lagrangian
\begin{eqnarray}
  \label{FP}
{\cal L}_{\rm FP} (x,y) &=& \bar{c}^a\, \frac{\delta F^a (A^a)}{\delta
\Theta^b}\, c^b \nonumber\\
&=& \bar{c}^a\,\Big[\, \partial^\mu\,
\big(\partial_\mu\delta^{ab}\: -\: g_5 f^{abc} A^c_\mu\,\big)\ -\
\xi\,\partial_5\,\big(\partial_5\delta^{ab}\: -\: g_5 f^{abc}
A^c_5\,\big)\,\Big]\,c^b\, .
\end{eqnarray}
In the above, $c^a  (x,y)$ denote the higher-dimensional ghost fields,
which are even under $Z_2$:  $c^a (x,y) = c^a(x,-y)$, i.e.~they  share
the same transformation properties with the group parameters $\Theta^a
(x,y)$.

In  Figs.~\ref{feynrules3gauge} and \ref{feynrules4gauge},  we exhibit
the Feynman rules for the  self-interactions of the KK modes $A^a_{(n)
\mu}$ and $A^a_{(n)  5}$ in the effective 4-dimensional  theory in the
\rxi-gauge   \equ{rxigaugefixtermnonab}.    In   the  unitary   gauge,
i.e.~$\xi  \to  \infty$, the  5D-QCD  Feynman  rules  reduce to  those
presented    in    \cite{DMN}.     The    factors    $\delta_{k,l,m}$,
$\tilde{\delta}_{l,k,m}$,            $\delta_{k,l,m,n}$            and
$\tilde{\delta}_{k,n,l,m}$  imply selection rules  for the  triple and
quartic coupling  of the KK  modes $A^a_{(n) \mu}$ and  $A^a_{(n) 5}$,
which  are typical  for the  interactions between  bulk  fields. These
factors are given by
\begin{equation}
\label{shorthandnotation}
\begin{split}
\delta_{k,l,m} & = \delta_{k+l+m,0} + \delta_{k+l-m,0} +
\delta_{k-l+m,0} + \delta_{k-l-m,0} \, , \\ \tilde{\delta}_{k,l,m} & =
- \delta_{k+l+m,0} + \delta_{k+l-m,0} - \delta_{k-l+m,0} +
\delta_{k-l-m,0} \, , \\
\end{split}
\end{equation}
for the triple gauge boson coupling and 
\begin{equation}
\label{shorthandnotation2}
\begin{split}
\delta_{k,l,m,n} & = + \delta_{k+l+m+n,0} + \delta_{k+l+m-n,0} +
                   \delta_{k+l-m+n,0} + \delta_{k+l-m-n,0} \\ & \quad
                   \, + \delta_{k-l+m+n,0} + \delta_{k-l+m-n,0} +
                   \delta_{k-l-m+n,0} + \delta_{k-l-m-n,0} \, , \\
                   \tilde{\delta}_{k,l,m,n} & = - \delta_{k+l+m+n,0} +
                   \delta_{k+l+m-n,0} + \delta_{k+l-m+n,0} -
                   \delta_{k+l-m-n,0} \\ & \quad \, -
                   \delta_{k-l+m+n,0} + \delta_{k-l+m-n,0} +
                   \delta_{k-l-m+n,0} - \delta_{k-l-m-n,0} \, ,
\end{split}
\end{equation}
for the quartic gauge boson coupling.

\setcounter{equation}{0}
\section{5-Dimensional Extensions of the Standard Model}
\label{MinimalExtensions}

In this section we shall study minimal 5-dimensional extensions of the
SM compactified  on an $S^1/Z_2$ orbifold, in  which the SU(2)$_L$ and
U(1)$_Y$  gauge bosons  as  well as the  Higgs  doublets  may not  all
propagate in the bulk.   In all these higher-dimensional scenarios, we
shall assume that the chiral fermions are  localized on a brane at the
$y=0$ fixed point of the $S^1/Z_2$ orbifold.

\subsection{SU(2)$_L \otimes $U(1)$_Y$-Bulk Model}
\label{allinthebulk}

To  start   with,   we  shall  first   consider    the most frequently
investigated model, where  all  electroweak gauge fields  propagate in
the bulk and couple to  both a brane  and a bulk  Higgs doublets.  The
Lagrangian   of  the  gauge-Higgs  sector  of  this higher-dimensional
standard model (HDSM) is given by
\begin{eqnarray}
\label{electroweaklagrangiangeneralhiggssector}
\la(x,y)  &=& - \, \frac{1}{4} \, B_{M N} \, B^{M N} \ - \ \frac{1}{4}
              \, F^a_{M N} F^{a M N} \ + \ \lr{D_M \, \Phi_1}\da \,
              \lr{D^M \, \Phi_1}\nonumber\\ 
&&+ \, \delta(y) \lr{D_{\mu} \,
              \Phi_2}\da \, \lr{D^{\mu} \, \Phi_2}\ - \ V(\Phi_1,
        \Phi_2) \ + \ \la_{\mr{GF}}(x,y)\ +\ \la_{\mr{FP}}(x,y) \, ,\quad
\end{eqnarray}
where $B_{M N}$ and $F^a_{M N}$ ($a = 1,2,3$  for SU(2)) are the field
strength    tensors   of the U(1)$_Y$   and   SU(2)$_L$  gauge fields,
respectively. As usual, we define the covariant derivative $D_M$ as
\begin{equation}
\label{covariantderivativesu2crossu1}
D_M\ = \ \partial_M \, - \, i \, g_5 \, A^a_M \, \tau^a \, - i \,
\frac{g_5'}{2} \, B_M\, .
\end{equation}
The Higgs  potential  $V(\Phi_1,  \Phi_2)$  of  this  SU(2)$_L \otimes
$U(1)$_Y$-bulk    model  has the       very same  analytic    form  as
in~\equ{generalhiggspotential}, where  $\Phi_1 (x,y)$ is  a bulk Higgs
doublet   and $\Phi_2 (x)$ a   brane  one.  After spontaneous symmetry
breaking,  the  Higgs doublets will  linearly be  expanded about their
VEVs, i.e.
\begin{equation}
\label{electroweakphiswithvevgeneralmodel}
\Phi_1(x,y) = 
\Vector{\chi_1^+ \\  
\frac{1}{\sqrt{2}} \, 
\Big(\,\frac{\displaystyle v_{1}}{\displaystyle \sqrt{2 \pi R}} \, + \, h_1 \, 
+ \, i \chi_1 \Big) } \, , \qquad
\Phi_2(x) =   
\Vector{\chi_2^+ \\  \frac{1}{\sqrt{2}} \, ( v_{2} \, + \, h_2 \, 
+ \, i \chi_2 ) } \, .
\end{equation}
Here, we shall not repeat the  calculational steps for determining the
particle mass spectrum  of the SU(2)$_L \otimes $U(1)$_Y$-bulk  model,
as they   are analogous to those   of the Abelian   model discussed in
Section~2.3 (see also Appendix~B).  In fact, the  above analogy in the
derivation of the particle   mass spectrum becomes rather  explicit if
the bulk gauge fields are written in terms of their higher-dimensional
mass eigenstates:
\begin{equation}
\label{su2u1states}
\begin{split}
W^{\pm}_M \, & = \, \frac{1}{\sqrt{2}} \, 
\lr{ A^1_M \, \mp \, i \, A^2_M } \, ,\\ 
Z_M \, & = \, \frac{1}{\sqrt{g_5^2 + \gs}} \, \lr{ g_5
\, A^3_M \, - \, g'_5 B_M } \, ,\\ 
A_M \, & = \, \frac{1}{\sqrt{g_5^2
+ \gs}} \, \lr{ g'_5 \, A^3_M \, + \, g_5 B_M } \, .
\end{split}
\end{equation}

Proceeding as in    the Abelian  case, we  may   easily  determine the
appropriate $R_\xi$ gauge-fixing   functions  for the SU(2)$_L$    and
U(1)$_Y$ gauge bosons:
\begin{eqnarray}
  \label{FaAa} 
F^a (A^a) &=& \partial_{\mu} A^{a \mu}\ -\ \xi_A\, \Bigl[
\partial_5 A^a_5\: -\: i \, \frac{g_5}{\sqrt{2 \pi R}} 
\lr{ \Phi_1\da \tau^a \Phi_0 - \Phi_0\da
\tau^a \Phi_1} \cob \nonumber\\ 
&& -\, i \, g_5 \lr{ \Phi_2\da \tau^a \Phi_0 - \Phi_0\da \tau^a 
\Phi_2 \! }\sib \, \delta(y) \Bigr] \, , \\
  \label{FB} 
F(B) & = & \partial_{\mu} B^{\mu}\ -\ \xi_B \Bigl[ \partial_5
B_5\: -\: i \, \frac{g'_5}{2 \, \sqrt{2 \pi R}} 
\lr{ \Phi_1\da \Phi_0 - \Phi_0\da \Phi_1} \cob\nonumber\\ 
&& -\, i \, \frac{g'_5}{2} \lr{ \Phi_2\da \Phi_0 - \Phi_0\da \Phi_2}
\sib \, \delta(y) \Bigr] \, ,
\end{eqnarray}
with
\begin{equation}
\label{defofphizero}
\Phi_0\ = \ \frac{1}{\sqrt{2}} \, \Vector{0 \\  v}\, ;\qquad
v   = \sqrt{v_{1}^2 +    v_{2}^2}\, .
\end{equation}
To avoid gauge-dependent photon-$Z$-mixing terms at the tree level, we
will assume in the following that it is always  $\xi_A = \xi_B = \xi$.
Under this assumption, the gauge-fixing  Lagrangian ${\cal L}_{\rm GF}
(x,y)$  in~\equ{electroweaklagrangiangeneralhiggssector}     may    be
expressed in terms of the real gauge-fixing  functions $F^a (A^a)$ and
$F(B)$ as follows:
\begin{equation}
  \label{LGFbulk}
{\cal L}_{\rm GF}(x,y) \ =\ - \, \frac{1}{2\xi}\, \big(F^a (A^a)\big)^2\
-\ \frac{1}{2\xi}\, \big(F (B)\big)^2\, .
\end{equation}
Furthermore,  the  Faddeev--Popov term    ${\cal   L}_{\rm FP}  (x,y)$
in~\equ{electroweaklagrangiangeneralhiggssector}  is induced  by   the
variations of  $F^a (A^a)$  and  $F(B)$ with respect  to SU(2)$_L$ and
U(1)$_Y$ gauge  transformations.  More explicitly, ${\cal  L}_{\rm FP}
(x,y)$ may be computed in the standard way by
\begin{equation}
{\cal L}_{\rm FP}(x,y)\ =\ \bar{c}^a\, \frac{\delta F^a (A^a)}{\delta
\Theta^b}\, c^b\ +\ \bar{c}\, \frac{\delta F (B)}{\delta
\Theta}\, c\, , 
\end{equation}
where $c^a(x,y)$ and   $c(x,y)$  are the   5-dimensional ghost  fields
associated     with   the  SU(2)$_L$   and    U(1)$_Y$   gauge groups,
respectively. As in the 5D-QCD, the ghost fields are even under~$Z_2$.

In the above  $R_\xi$-gauge-fixing prescription, the complete  kinetic
Lagrangian of the gauge sector written in terms  of the fields defined
in \equ{su2u1states} becomes rather analogous to the corresponding one
of the   Abelian  model  investigated in   Section~\ref{twoHiggs}.  In
Appendix~B,   we give the  propagators  of the KK  gauge and Goldstone
modes in  the $R_\xi$ gauge,together  with  the exact analytic results
for the couplings of  the gauge bosons to fermions  to be discussed in
Section~\ref{sectiononfermion}.

\subsection{SU(2)$_L$-Brane, U(1)$_Y$-Bulk Model}
\label{su2braneu1bulk}

Let us now consider a new minimal 5-dimensional alternative to the SM,
in which only  the U(1)$_Y$ gauge boson propagates  in the bulk, while
the SU(2)$_L$ gauge field lives on the $y=0$ boundary of the $S^1/Z_2$
orbifold.   The  Lagrangian  of  this  SU(2)$_L$-brane,  U(1)$_Y$-bulk
model~is
\begin{eqnarray}
\label{electroweaklagrangianuoneinbulk}
\la (x,y) &=& - \, \frac{1}{4} \, B_{M N} \, B^{M N} \ + \ \delta(y) \,
              \Big[\, - \, \frac{1}{4} \, F^a_{\mu \nu} F^{a \mu \nu}\:
	      + \: \lr{D_{\mu} \, \Phi}\da \, \lr{D^{\mu} \, \Phi}\: - 
	      \: V(\Phi)\, \Big]\nonumber\\
&&+ \, \la_{\mr{GF}}(x,y)\ +\ \la_{\mr{FP}}(x,y) \, .
\end{eqnarray}
Observe that only a brane Higgs doublet
\begin{equation}
\Phi(x) \, = \, 
\Vector{\chi_2^+ \\ \frac{1}{\sqrt{2}} \, ( v \, + \, h \, + \, i \chi ) }
\end{equation}
can  be added in this  model. The reason is that  a bulk Higgs doublet
would destroy the gauge  invariance of the theory  in the bulk if  one
coupled it to the covariant derivative $D_{\mu} \, = \, \partial_{\mu}
\, - \, i \, g \,  A^a_{\mu}(x) \, \tau^a \, -  i \, \frac{g_5'}{2} \,
B_{\mu}(x,0)$  on  the  $y=0$ brane.    As  a consequence,   the Higgs
potential of this model has the known SM form: $V(\Phi) = \mu^2 | \Phi
|^2 + \lambda | \Phi |^4$.

In the SU(2)$_L$-brane, U(1)$_Y$-bulk  model, only the  $B^{\mu}(x,y)$
boson has  to  be expanded in  Fourier  modes.  Although the $W$-boson
sector   is   completely  standard,  the   neutral   gauge sector gets
complicated by the brane-bulk  mixing of  $B_\mu (x,y)$ with  $A^3_\mu
(x)$ through the VEV of the brane Higgs field  $\Phi (x)$.  To be more
precise, we find the  effective mass-matrix Lagrangian of the  neutral
gauge sector
\begin{equation}
\la^N_{\rm mass} (x)\ =\ \frac{1}{2} \left( A^{3 \, \mu} , \,
B_{(0)}^{\mu} , \, B_{(1)}^{\mu} , \, \ldots \right) M_{N}^2 \left(
\begin{array}{c}
A^{3}_{\mu} \\ 
B_{(0) \mu} \\
B_{(1) \mu} \\
\vdots
\end{array}
\right)
\end{equation}
with
\begin{equation}
  \label{MNY}
M_N^2 \ = \ 
 \left(
\begin{array}{cccc}
m^2 \frac{g^2}{g'{}^2} & - m^2 \frac{g}{g'} & - \sqrt{2} \, m^2
\frac{g}{g'} & \cdots \\ - m^2 \frac{g}{g'} & m^2 & \sqrt{2} \, m^2 &
\cdots \\ - \sqrt{2} \, m^2 \frac{g}{g'} & \sqrt{2} \, m^2 & 2 \, m^2 \,
+ \, (1/R)^2 & \cdots \\ \vdots & \vdots & \vdots & \ddots
\end{array} \right)
\end{equation}
and $g'_5 = g' \sqrt{2 \pi R}$, $m^2 =  g'{}^2 v^2/4$. The mass matrix
$M^2_N$ contains a  zero  eigenvalue  which  corresponds to a massless 
photon $\hat{A}_\mu$, i.e.
\begin{equation}
\hat{A}_\mu \ =\ s_w\, A^3_\mu \, + \, c_w\, B_{(0)\mu}\, ,
\end{equation}
where $s_w  = \sqrt{1-c^2_w}  = g'/\sqrt{g^2  +  g^{\prime 2}}$ is the
sine of  the weak mixing  angle.  The other  non-zero mass eigenvalues
$m_{Z (n)}$ of $M^2_N$ in~\equ{MNY} may  be determined by the roots of
the transcendental equation
\begin{equation}
m_{Z (n)} \ = \ \pi \, m^2 \, R \, \cot \lr{\pi \, m_{Z (n)} \, R}
\ + \ \frac{g^2}{g'{}^2} \, \frac{m^2}{m_{Z (n)}} \, .
\end{equation}
The respective mass eigenstates are given by  
\begin{equation}
\hat{Z}^{\mu}_{(n)} \ = \ \frac{1}{N} \, \bigg[ \frac{m_Z}{m_{Z
(n)}} \, c_w \, A^{3  \,  \mu} \  
-\ \, \sum^{\infty}_{j=0} \, 
\frac{ \sqrt{2} \,  m_{Z  (n)} \, m_Z}{m^2_{Z(n)} \,  -  \, (j / R)^2}
\, \lr{\frac{1}{\sqrt{2}}}^{\delta_{j,0}} \, s_w\, B^{\mu}_{(j)}\, \bigg] \, ,
\end{equation}
where $m_Z =  \sqrt{g^2 + g'{}^2}\, v/2$,
\begin{equation}
\label{normalizationbranebulk}
N^2 \ = \ \frac{1}{2} \, 
\bigg[\,\frac{c^2_w}{s^2_w} 
\bigg(\,\frac{m_Z^2}{m^2_{Z(n)}}\: -\: 2\bigg)\ +\ s^2_w \,
\pi^2  m_Z^2 R^2\: +\: \frac{m_{Z (n)}^2}{m_Z^2s^2_w}\: +\: 1\,
\bigg]\, .
\end{equation}
Notice that the KK mass eigenmode $\hat{Z}_{(0)}$ has to be identified
with the observable $Z$ boson.

In analogy to the  SM-bulk model, the appropriate $R_\xi$ gauge-fixing
functions for this brane-bulk model are written
\begin{eqnarray}
  \label{FaAabranebulk} 
F^a (A^a) &=& \partial_{\mu} A^{a \mu}\ +\ \xi\, 
i g \lr{ \Phi\da \tau^a \Phi_0 - \Phi_0\da
\tau^a \Phi}\\ 
  \label{FBbranebulk} 
F(B) & = & \partial_{\mu} B^{\mu}\ -\ \xi \Bigl[ \partial_5
B_5\: -\, i \frac{g'_5}{2} \lr{ \Phi\da \Phi_0 - \Phi_0\da \Phi} 
\delta(y) \Bigr] \, ,
\end{eqnarray}
with
\begin{equation}
\label{defofphizerobranebulk}
\Phi_0\ = \ \frac{1}{\sqrt{2}} \, \Vector{0 \\  v}\, .
\end{equation}
Nevertheless, because of    the specific brane-bulk   structure of the
higher-dimensional  model,  the corresponding  gauge-fixing Lagrangian
has now the form
\begin{equation}
  \label{LGFbranebulk}
{\cal L}_{\rm GF}(x,y) \ =\ - \, \frac{1}{2\xi}\, 
\big(F^a (A^a)\big)^2\ \, \delta(y) \  -\ 
\frac{1}{2\xi}\, \big(F (B)\big)^2\, .
\end{equation}
Like the charged gauge sector, the charged scalar sector is completely
standard  in  this model.  The neutral  scalar  sector, however, has a
structure very  similar to the one  of the  Abelian model discussed in
Section~2.2.  Again, one  can  show  the  existence of  an  one-to-one
correspondence  between  the KK gauge modes  with mass $m_{Z (n)}$ and 
their associate  would-be   Goldstone   modes  with   mass $\sqrt{\xi}
m_{Z (n)}$.  The latter KK modes are given by
\begin{equation}
\label{wouldbeGoldstonemasseigenstatesbranebulk}
\hat{G}^0_{(n)}\ =\ \frac{1}{N}
\, \bigg(\, \chi \ -\ \frac{g'v}{\sqrt{2}}\,
\sum^{\infty}_{j=1} \, 
\frac{j/R}{m^2_{Z(n)} \, - \, (j/R)^2} \, B_{(j) 5}\,\bigg) \, , 
\end{equation}
where      the     normalization     factor     $N$     is     defined
in~\equ{normalizationbranebulk}.

The Faddeev-Popov Lagrangian ${\cal L}_{\rm FP}$  can also be obtained
in the standard fashion. Taking  the brane-bulk structure of the model
into account, we may determine ${\cal L}_{\rm FP}$ by
\begin{equation}
  \label{FPbrane}
{\cal L}_{\rm FP}(x,y)\ =\ \bar{c}^a(x)\, 
\frac{\delta F^a (A^a(x))}{\delta
\Theta^b(x)}\, c^b(x)\, \delta(y)\ +\ \bar{c}(x,y)\, 
\frac{\delta F (B(x,y))}{\delta \Theta(x,y)}\, c(x,y)\, ,
\end{equation}
where the $(x,y)$-dependence  of the different quantities involved  is
explicitly indicated.

\subsection{SU(2)$_L$-Bulk, U(1)$_Y$-Brane Model}
\label{branebulk}

Another minimal  5-dimensional extension  of the SM,  complementary to
the one discussed in Section~4.2, emerges if the SU(2)$_L$ gauge boson
is  the only  field  that  feels  the presence of   the  fifth compact
dimension. By analogy, the Lagrangian of this model reads
\begin{eqnarray}
\label{electroweaklagrangiansutwoinbulk}
\la (x,y) &=& - \, \frac{1}{4} \, F^a_{M N} F^{a M N} \  + \ \delta(y)
 \, \Big[\, - \, \frac{1}{4} \, B_{\mu \nu} \, B^{\mu \nu}\: + \: 
 \lr{D_{\mu} \, \Phi}\da \, \lr{D^{\mu} \, \Phi}\: - \: V(\Phi)\, 
 \Big]\nonumber\\
&&+ \, \la_{\mr{GF}}(x,y)\ +\ \la_{\rm FP}(x,y)\, ,
\end{eqnarray}
with $D_{\mu} \, = \, \partial_{\mu} \, -\, i \, g_5 \, A^a_{\mu}(x,0)
\, \tau^a \, -   i \, \frac{g'}{2}  \, B_{\mu}(x)$.   As in  the model
discussed in the  previous section, there is only  one  Higgs field on
the brane $y=0$  and the Higgs potential is  of the  SM form.  Because
only the SU(2)$_L$ gauge  boson lives in the  bulk, the charged  gauge
sector of this higher-dimensional standard model is equivalent to that
of the SM-bulk model discussed in Section~4.1 in the limit $\sin \beta
\to 0$, i.e.~only  the  Higgs  field  restricted to the  brane  $y =0$
acquires  a    non-vanishing    VEV.     Thus,    the  SU(2)$_L$-bulk,
U(1)$_Y$-brane model   predicts a KK   tower of $W$-boson excitations,
while the neutral gauge sector is quite analogous to the one discussed
in  the  previous section.   Specifically, the  effective  mass-matrix
Lagrangian of the neutral gauge sector is given by
\begin{equation}
\la^N_{\rm mass}(x)\ =\ 
\frac{1}{2}\, 
\left( B^{\mu} , \, A^{3 \, \mu}_{(0)} , \, A^{3 \, \mu}_{(1)} , \, 
\ldots \right)
M_N^2
\left(
\begin{array}{c}
B_{\mu} \\
A^{3}_{(0) \mu} \\ 
A^{3}_{(1) \mu} \\ 
\vdots
\end{array}
\right),
\end{equation}
with
\begin{equation}
M_N^2 \, = \,  \left(
\begin{array}{cccc}
m^2 \frac{g'{}^2}{g^2} & - m^2 \frac{g'}{g} & - \sqrt{2} \, m^2
\frac{g'}{g} & \cdots \\ - m^2 \frac{g'}{g} & m^2 & \sqrt{2} \, m^2 &
\cdots \\ - \sqrt{2} \, m^2 \frac{g'}{g} & \sqrt{2} \, m^2 & 2 \, m^2 \,
+ \, (1/R)^2 & \cdots \\ \vdots & \vdots & \vdots & \ddots
\end{array}\right), 
\end{equation}
$g_5=g \sqrt{2 \pi R}$ and  $m^2 = g^2v^2/4$. Again,  we find that the
zero  KK  mode  given by  the  linear  combination $\hat{A}_\mu  = s_w
A^3_{(0)\mu}   + c_w B_\mu$   represents a massless  vector field, the
photon. The higher   KK modes  are  massive  and their  masses  may be
obtained by the solutions of the transcendental equation
\begin{equation}
m_{Z (n)} \ = \ \pi \, m^2 \, R \, \cot \lr{\pi \, m_{Z (n)} \, R}
\ + \ \frac{g'{}^2}{g^2} \, \frac{m^2}{m_{Z (n)}} \,  .
\end{equation}
The $Z$   boson, denoted as     $Z_{(0)}$, and  its heavier  KK   mass
eigenmodes   may be  conveniently expressed   in terms   of  the gauge
eigenstates as
\begin{equation}
\hat{Z}^{\mu}_{(n)} \ = \ \frac{1}{N} \, \bigg[\, \sum^{\infty}_{j=0} \, 
\frac{\sqrt{2} \, m_{Z (n)} \, m_Z}{m^2_{Z(n)} \, - \,
(j / R)^2} \, \bigg(\,\frac{1}{\sqrt{2}}\,\bigg)^{\delta_{j,0}} \, 
c_w \, A^{3 \, \mu}_{(j)} \ 
- \ \frac{m_Z}{m_{Z (n)}} \, s_w \, B^{\mu}\, 
\bigg] \, ,
\end{equation}
where
\begin{equation}
\label{normalizationbulkbrane}
N^2 \ = \ \frac{1}{2} 
\bigg[\, \frac{s^2_w}{c^2_w} 
\bigg(\,\frac{m_Z^2}{m^2_{Z(n)}}\: -\: 2\bigg)\ +\ 
c^2_w \, \pi^2 m_Z^2 R^2\: +\: 
\frac{m_{Z(n)}^2}{m_Z^2 c^2_w} \: +\: 1\, \bigg] \, .
\end{equation}

In close analogy    to the previous  section, the   higher-dimensional
gauge-fixing  functions leading to  the  generalized $R_\xi$-gauge are
given by
\begin{eqnarray}
  \label{FaAabulkbrane} 
F^a (A^a) &=& \partial_{\mu} A^{a \mu}\ -\ \xi\, \Bigl[ \partial_5
A^a_5\: -\, i g_5 \lr{ \Phi\da \tau^a \Phi_0 - \Phi_0\da
\tau^a \Phi} \delta(y) \Bigr] \, ,\\ 
  \label{FBbulkbrane} 
F(B) & = & \partial_{\mu} B^{\mu}\ +\ \xi \, 
i \, \frac{g'}{2} \lr{ \Phi\da \Phi_0 - \Phi_0\da \Phi} \, ,
\end{eqnarray}
giving rise to  the gauge-fixing  Lagrangian 
\begin{equation}
  \label{LGFbulkbrane}
{\cal L}_{\rm GF}(x,y) \ =\ \, - \, \frac{1}{2\xi}\, 
\big(F^a (A^a)\big)^2\ \,  -\ 
\frac{1}{2\xi}\, \big(F (B)\big)^2\, \delta(y) \, .
\end{equation}
The charged  scalar sector of  this model is  identical to that of the
SM-bulk model of Section~4.1, without the presence of a Higgs field on
the $y=0$  boundary.   On the  other hand,  the  neutral scalar sector
predicts a  KK tower of would-be  Goldstone modes  associated with the
longitudinal  polarization degrees  of   the massive   KK gauge  modes
$\hat{Z}_{(n)}$.  The would-be KK Goldstone modes are  determined 
by
\begin{equation}
\label{wouldbeGoldstonemasseigenstatesbulkbrane}
\hat{G}^0_{(n)}\ =\ \frac{1}{N}
\, \bigg(\, \chi \ +\ 
\frac{gv}{\sqrt{2}}\,
\sum^{\infty}_{j=1} \, \frac{j/R}{m^2_{Z(n)} \, - \, (j/R)^2}\, 
A^3_{(j) 5}\, \bigg) \, ,
\end{equation}
with $N$ as defined in \equ{normalizationbulkbrane}. The Faddeev-Popov
Lagrangian   can be calculated  as  in the  model described earlier in
Section~4.2  (cf.~\equ{FPbrane}),    by   considering    the   obvious
modifications that  take  account  of  the complementary    brane-bulk
structure of the model.

\subsection{Localization of Fermions on the Brane}
\label{sectiononfermion}

In the   minimal  5-dimensional extensions   of the SM  we  have  been
studying, we have assumed  that all the SM  fermions are  localized at
the  $y=0$ fixed  point of the  $S^1/Z_2$  orbifold.  Therefore,  upon
integrating out the $y$ dimension, both the effective kinetic terms of
fermions  and   the  effective Yukawa   sector  take  on   the   usual
4-dimensional SM structure.  Clearly, the  SM fermions do not have  KK
modes.    Under a gauge transformation,  the  left-  and right- handed
fermions transform according to
\begin{equation}
\begin{split}
\Psi_L(x) \quad & \to \quad \exp 
\Big(ig_5 \, \Theta^a(x,0) \, \tau^a\, + \, ig_5' \, Y^L \, \Theta (x,0)
\Big)\, \Psi_L(x) \, , \\
\Psi_R(x) \quad & \to \quad \exp \Big( ig_5' \, Y^R \, \Theta (x,0)
\Big)\, \Psi_R(x)\, .
\end{split}
\end{equation}
The corresponding    covariant derivatives  that   couple  the  chiral
fermions to the gauge fields are given by
\begin{equation}
\begin{split}
D^L_{\mu} \, & = \, \partial_{\mu} \, - \, i \, g_5 \, A^a_{\mu}
\, \tau^a \, - \, i \, g_5' \, Y^L \, B_{\mu} \, ,\\ 
D^R_{\mu} \, & = \, \partial_{\mu} \, - \, i \, g_5'\, Y^R \, 
B_{\mu} \, .
\end{split}
\end{equation}

It  is obvious that the effective   coupling of a  fermion  to a gauge
boson restricted to  the same brane $y=0$  has  its SM value.  On  the
other hand,   the   effective interaction Lagrangian  describing   the
coupling of  a fermion to a  gauge boson  living  in the  bulk has the
generic form
\begin{equation}
\label{interactionsfermwithgbmodesabelianmodel}
\la_{\rm int}(x)\ = \ \Psibar\, \gamma^{\mu} \, \lr{g_V
                        + g_A \gamma^5} \, \Psi\, \Big( A_{(0)
                        \mu} \, + \, \sqrt{2} \, \sum_{n=1}^{\infty}
                        A_{(n) \mu} \Big) \, .
\end{equation}
Again, the coupling parameters $g_V$ and $g_A$ are  set by the quantum
numbers   of the zero   KK  gauge mode  and   receive their SM values.
Because the KK mass eigenmodes    generally differ from the    Fourier
modes,   their couplings to fermions $g_{V(n)}$ and $g_{A(n)}$ have to 
be calculated for  each model individually by  taking into account the 
appropriate  weak-basis  transformations.      The precise  values  of 
$g_{V(n)}$     and     $g_{A(n)}$   will  be  very  important for  our 
phenomenological discussion in the next section. The Feynman rules for
the  interactions  of  the  KK  gauge  mass eigenmodes to fermions are 
exhibited in Appendix~B.

Likewise, the Feynman rules for the interaction of the Goldstone modes
to fermions can also be obtained from the SM Yukawa sector by relating
the KK weak  modes to the respective  KK mass eigenmodes.  It is worth
remarking  here that although the $Z_2$-odd  fifth component of a bulk
gauge boson  $A_M$,  $A_5$,  does  not  couple directly to   the brane
fermions, $A_5$ is  involved in fermionic couplings  due to its mixing
with   the CP-odd  Higgs    fields which are   even  under  $Z_2$.  In
particular, one can  show that  the  resulting Goldstone  couplings to
fermions have the proper analytic structure to assure gauge invariance
in the computation of S-matrix elements.

\setcounter{equation}{0}
\section{Global-Fit Analysis}
\label{Phenomenology}

In this section, we shall  evaluate the bounds on the compactification
scale $1/R$  of  minimal higher-dimensional  extensions  of the SM  by
analyzing a  large  number of high  precision electroweak observables.
To be  specific, we proceed as follows.   We  relate the SM prediction
$\mathcal{O}^{\mr{SM}}$   for   an  electroweak  observable   to   the
prediction $\mathcal{O}^{\mr{HDSM}}$ for  the same observable obtained
in the higher-dimensional SM under investigation through
\begin{equation}
\label{generalformofpredictions}
{\cal O}^{\rm HDSM} \ =\ {\cal O}^{\rm SM} \, \big( 1\: +\: 
\Delta^{\rm HDSM}_{\cal O} \big)\, .
\end{equation}
Here, $\Delta^{\rm HDSM}_{\cal O}$ is the tree-level modification of a
given observable ${\cal O}$  from its SM  value due to the presence of
one extra dimension.  To enable a direct comparison of our predictions
with  the    electroweak  precision data~\cite{PDG},   we  include  SM
radiative corrections to~$\mathcal{O}^{\mr{SM}}$.  However, we neglect
SM- as well as KK- loop  contributions to $\Delta^{\rm HDSM}_{\cal O}$
as higher order effects.

As input SM parameters for  our theoretical predictions, we choose the
most accurately measured ones,  namely  the $Z$-boson mass $M_Z$,  the
electromagnetic  fine structure   constant~$\alpha$  and   the   Fermi
constant  $G_F$.  In all the   5-dimensional models  under study,  the
tree-level    $Z$-boson   mass $m_{Z (0)}$ generally deviates from its
SM  form $m_Z = \sqrt{g^2 + g'{}^2}\, v/2$. Therefore, we parameterize
this deviation as follows:
\begin{equation}
\label{deltazdef}
m^2_{Z (0)} \  =\ m^2_Z \, \lr{1 \: + \: \Delta_{Z} X} \, ,
\end{equation}
where 
\begin{equation}
  \label{X}
X\ =\ \frac{\pi^2}{3} \, \frac{m^2_Z}{M^2}\ 
\end{equation}
(with  $M   = 1/R$)  represents  the  typical  scale  quantifying  the
higher-dimensional  effect   and   $\Delta_{Z}$ is  a  model-dependent
parameter of order  unity.  Since the massless  photon retains its  SM
properties   through  the  entire   process  of compactification,  the
electromagnetic fine structure constant is still given by its SM value
\begin{equation}
\alpha\ =\ \frac{e^2}{4 \pi} \, .
\end{equation}
Instead,  the Fermi constant $G_F$  as determined by the muon lifetime
may receive direct as well as indirect contributions due to KK states.
We may account for this modification of $G_F$ by writing
\begin{equation}
\label{thetaandinput}
G_F \ = \ \frac{\pi \alpha}{\sqrt{2} s^2_w \, c^2_w \, m_{Z (0)}^2} \,
\lr{ 1\: +\: \Delta_G X} \, ,
\end{equation}
where the order unity  parameter   $\Delta_G$ strongly depends on  the
details of the 5-dimensional model under consideration.

In the computation  of the electroweak  precision observables, it will
be necessary to express  the weak mixing  angle $\theta_w$ in terms of
the  input  parameters  $\alpha$, $m_{Z (0)}^2$ and $G_F$, by means of 
\equ{thetaandinput}. In this  respect,  it  is  useful  to  define  an
effective  weak mixing angle $\hat{\theta}_w$  using the tree-level SM
relation
\begin{equation}
\label{definitionequationfonormalizedthetaw}
G_F \ = \ \frac{\pi \alpha}{\sqrt{2} \hat{s}^2_w \, \hat{c}^2_w \,
m_{Z (0)}^2} \ .
\end{equation}
With     the     above   definition     for    $\hat{\theta}_w$    and
\equ{thetaandinput},  we may relate the  squared sines of the two weak
mixing angles by
\begin{equation}
\label{thetawhatthroughthetaw}
\hat{s}^2_w \ = \ s^2_w \, \lr{ 1\: +\: \Delta_{\theta} X} \, .
\end{equation}
Again,     $\Delta_{\theta}$     in~\equ{thetawhatthroughthetaw} is  a
model-dependent parameter of order unity to be determined below.

\subsection{SU(2)$_L \otimes $U(1)$_Y$-Bulk Model}

Before we present predictions  for the electroweak observables  in the
SU(2)$_L  \otimes  $U(1)$_Y$-bulk model, let   us   first give the  KK
modifications   for   some  of  the  fundamental   parameters  of  the
theory.  The KK modifications of the   $Z$- and $W$-  boson masses are
found to be
\begin{equation}
\label{normalizedmassesthroughbaremasses}
\begin{split}
\Delta_Z \, & = \, - \, s^4_{\beta} \, , \\
\Delta_W \, & = \, - \, s^4_{\beta} \, \hat{c}^2_w \, , 
\end{split}
\end{equation}
where $\Delta_W$  is   defined in  analogy    to \equ{deltazdef}.   In
\equ{normalizedmassesthroughbaremasses}  and in the following, we will
often use the    following  short-hand notations   for   trigonometric
functions: $s_x = \sin x$, $c_x = \cos x$, $s_{2x} = \sin 2x$, $c_{2x}
=  \cos 2x$. 

KK effects also  cause tree-level shifts to   the $W$- and $Z$-  boson
gauge couplings.  The physical gauge-boson couplings are given by
\begin{equation}
\label{normalizedcouplingsthroughbarecoupings}
\begin{split}
g_{W (0)} \ & = \ g \, \lr{ 1\: -\:
s^2_{\beta} \, \hat{c}^2_w\, X} \, , \\ g_{Z (0)}
\ & =\ g \, \lr{ 1\: -\: s^2_{\beta} \, X } \, .
\end{split}
\end{equation}
These last two relations are  approximate,  i.e.~they are obtained  by
expanding  the exact  analytic  results for  the masses and couplings,
stated in Appendix~B,  to leading order in  the parameter  $X$ defined
in~\equ{X}.  Finally, the KK tree-level  shift $\Delta_G$ of the Fermi
constant $G_F$ is
\begin{equation}
\Delta_G \ = \ \hat{c}^2_w \lr{1\: -\: 2 s^2_{\beta}\: -\:
\frac{\hat{s}^2_w}{\hat{c}^2_w} s^4_{\beta}}\ ,
\end{equation}
which implies
\begin{equation}
\Delta_{\theta}  \  = \ -\, \frac{\hat{c}^4_w}{\hat{c}_{2  w}} \lr{1\:
-\: 2   s^2_{\beta}\: -\: \frac{\hat{s}^2_w}{\hat{c}^2_w} s^4_{\beta}}
\, .
\end{equation}
Notice that $\Delta_\theta$ determining  the difference  between $s_w$
and  $\hat{s}_w$ is   a  key  parameter in   the  computation  of many
precision observables, as   it  additionally  enters  via the   vector
coupling of the $Z$ boson.

In our calculations of the electroweak observables to leading order in
$X$, we consistently   use  $m_{Z (n)} \approx  m_{W (n)} \approx n/R$
and $g_{Z (n)}  \approx  g_{W (n)} \approx \sqrt{2} g$ for $n \ge  1$.
Within  this  approximative  framework,  we compute the following high  
precision  observables:   the  $W$-boson  mass $M_W$, the    $Z$-boson 
invisible  width  $\Gamma_Z (\nu \ov{\nu})$, $Z$-boson leptonic widths
$\Gamma_Z(l^+l^-)$, the $Z$-boson hadronic width $\Gamma_Z(\mr{had})$,
the  weak  charge of cesium $Q_W$ measuring atomic   parity violation, 
various ratios $R_l$ and $R_q$ involving partial $Z$-boson  widths and
the $Z$-boson hadronic width, fermionic asymmetries $A_f$  at  the $Z$
pole,       and    various  fermionic forward-backward     asymmetries  
$A_{\mr{FB}}^{(0,f)}$  at  vanishing  polarization.  A  complete  list 
of  the  considered  observables  along  with  the  SM predictions and 
their   experimental values is  given in Appendix~C.

\begin{table}[t]
\renewcommand{\arraystretch}{1.5}
\begin{center}
\begin{tabular}{c|c}
\hline \hline
Observable & $\Delta^{\mr{HDSM}}_{\mathcal{O}}/X$ \\ \hline \hline
$M_W$ & $\frac{1}{2} \lr{s^4_{\beta} \, \, \hat{s}^2_w \, + \,
\frac{\hat{s}^2_w}{\hat{c}^2_{w}} \, \Delta_{\theta}}$ \\[0.75ex]
\hline $\Gamma_Z (\nu \ov{\nu})$ & $\hat{s}^2_w \, \lr{s^2_{\beta} -
1}^2 - 1$ \\ \hline $\Gamma_Z (l^+ l^-)$ & $\hat{s}^2_w \,
\lr{s^2_{\beta} - 1}^2 \, - \, 1 \, + \, \Delta_l$ \\ \hline $\Gamma_Z
(\mr{had})$ & $\hat{s}^2_w \, \lr{s^2_{\beta} - 1}^2 \, - \, 1 \, + \,
\Delta_h$ \\ \hline $Q_W (\mr{Cs})$ & $\lreckig{ \lr{ \! 1 -
s^2_{\beta} \! }^2 \, + 4 \, Z \, \lr{ Q_W^{\mr{SM}}}^{\! -1} \,
\Delta_{\theta} } \, \hat{s}^2_w $ \\[0.75ex] \hline $R_{l}$ & $ - \,
\Delta_{l} \, + \, \Delta_h$ \\ \hline $R_{q}$ & $ \Delta_{q} \, - \,
\Delta_h$ \\ \hline $A_{f}$ & $\Delta_V \, - \, \Delta_{f}$ \\
[0.75ex] \hline $A_{\mr{FB}}^{(0,f)}$ & $\Delta_V \, - \, \Delta_{f}
\, + f \leftrightarrow e$ \\ [0.75ex] 
\hline\hline
\end{tabular}
\end{center}
\mycaption{\label{tab1} Predictions for
$\Delta^{\mr{HDSM}}_{\mathcal{O}}/X$ in the SU(2)$_L \otimes
$U(1)$_Y$-bulk model. The auxiliary parameters $\Delta_V$, $\Delta_f$
and $\Delta_h$ are defined in \equ{deltadefinitions}.}
\end{table}

In   Table~\ref{tab1},  we  present   predictions for   the  parameter
$\Delta^{\mr{HDSM}}_{\mathcal{O}}/X$  in    the SM-bulk  model,  where
$\Delta^{\mr{HDSM}}_{\mathcal{O}}$   and      $X$ are     defined   by
\equ{generalformofpredictions}  and \equ{X}, respectively.   Moreover,
the auxiliary parameters that occur in Table~\ref{tab1} are given by
\begin{equation}
\label{deltadefinitions}
\begin{split}
\Delta_V \ & = \ \frac{4 \, Q_{f} \, \hat{s}^2_w}{2 T_{3 f} - 4
Q_{f} \hat{s}^2_w} \: \Delta_{\theta}\,, \\[2ex] 
\Delta_f \ & = \  \frac{8 \,
\hat{s}^2_w \, Q_f \, (2 T_{3f} - 4 Q_f \hat{s}^2_w)}{(2 T_{3f} -
4 Q_f \hat{s}^2_w)^2 + (2 T_{3f})^2} \: \Delta_{\theta}\,, \\[2ex]
\Delta_h \ & = \  \frac{8 \, \hat{s}^2_w \, \sum_{q} \, Q_q
\, \lr{2 T_{3q} - 4 Q_q \hat{s}^2_w} } {\sum_{q} \lreckig{(2
T_{3q} - 4 Q_q \hat{s}^2_w)^2 + (2 T_{3q})^2} } \: \Delta_{\theta} 
\,,\\[2ex] Q_W^{\mr{SM}}\ &=\  \lr{Z - N}\ -\ 4 \, Z\, \hat{s}^2_w\,,
\end{split}
\end{equation}
where   $Q_f$ and  $T_{3f}$  are  the  electric charge  and  the third
component   of   the   weak   isospin  of a fermion $f$, respectively, 
$q=u,d,c,s,b$  and  $N = 78$  is the number of neutrons and $Z=55$ the  
number  of  protons in the cesium nucleus.  In~\equ{deltadefinitions}, 
the  parameters  $\Delta_V$,  $\Delta_f$  and   $\Delta_h$   are   all 
proportional  to $\Delta_\theta$, since  they arise from  substituting
$s^2_w$ by $\hat{s}^2_w$ into the different  electroweak  observables.
In detail,   $\Delta_V$  parameterizes  the  KK  shift  in  the vector  
coupling  of  the  $Z$  boson  to fermions. $\Delta_f$ results from an 
analogous  KK  shift in the sum of the squared vector and axial vector 
couplings for a given fermion $f$. Similarly, $\Delta_h$ gives  the KK
shift  in the total hadronic width of the $Z$ boson.

Employing   the  results  of     $\Delta^{\mr{HDSM}}_{\mathcal{O}}$ in
Table~\ref{tab1}, we can     compute  the predictions for  all     the
electroweak    observables    listed  in    Appendix~C,    by   virtue
of~\equ{generalformofpredictions}.  We will confront these predictions
with  the respective experimental   values, which are  also listed  in
Appendix~C. To do so,  we perform a  $\chi^2$ test to obtain bounds on
the  compactification scale $M=1/R$  as  a function  of the bulk-brane
angle $\sin\beta$.   Thus,    in our  global-fit   analysis  (ignoring
correlation effects between the observables to first approximation), a
compactification   radius  is considered to   be  compatible at the $n
\sigma$ confidence  level (CL), if $\chi^2  (R) -  \chi^2_{\mr{min}} <
n^2$, where
\begin{equation}
  \label{chi2}
\chi^2(R) \ = \ \sum_i \,\frac{\lr{ \mathcal{O}_i^{\mr{exp}}\: -\:
\mathcal{O}_i^{\mr{HDSM}} }^2}{\lr{ \Delta \mathcal{O}_i}^2}
\end{equation}
and    $\chi^2_{\mr{min}}$   is  the  minimum   of    $\chi^2$ for   a
compactification  radius in  the physical   region, i.e.~for  $R^2>0$.
In~\equ{chi2},   $i$   runs  over  all   the   observables  listed  in
Table~\ref{observablevaluesandsmpredictions}  in  Appendix~C.   {}From
this   table,  one easily    sees  that  the total   experimental  and
theoretical uncertainty $\lr{\Delta \mathcal{O}_i}^2$ of an observable
${\cal O}_i$ is dominated by its experimental uncertainty. 

\begin{figure}[t]
\begin{center}

\includegraphics{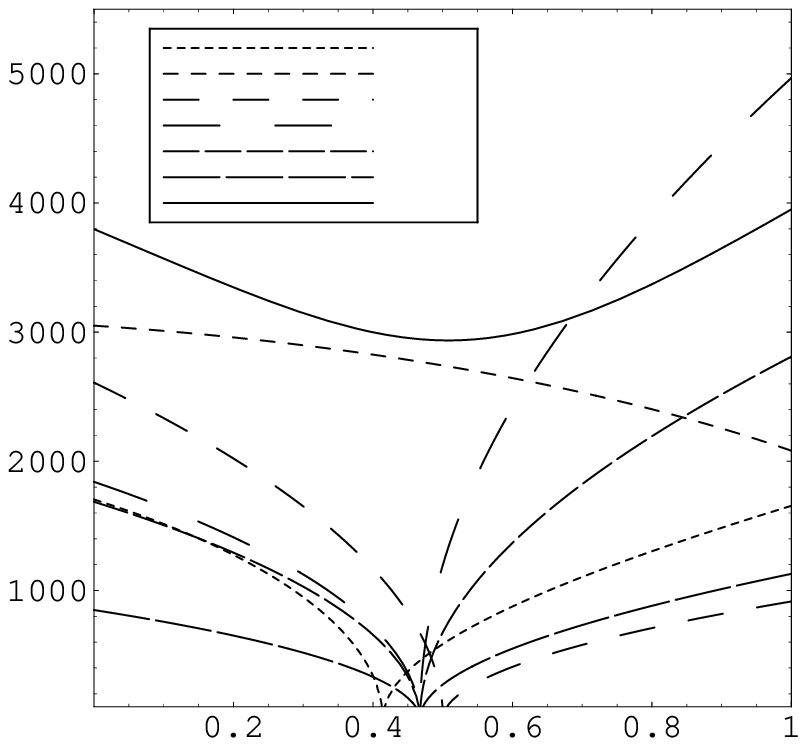}

\begin{picture}(0,0)(0,0)	
\SetWidth{1.}
\thicklines

\put(3,0){$\sin^2 \beta$} 
\put(-130,55){\begin{rotate}{90} Compactification scale $1/R$ [GeV] 
\end{rotate}} 

\put(-6,220){\tiny $M_{W}$}
\put(-6,213){\tiny $\Gamma_{Z}\!({\rm had})$}
\put(-6,205){\tiny $R_{\tau}$}
\put(-6,196){\tiny $A_{\mr{FB}}^{(0,b)}$}
\put(-6,189){\tiny $A_{\tau}$}
\put(-6,183){\tiny $Q_{W}$}
\put(-6,175){\tiny $\chi^2$}

\end{picture}

\end{center}
\mycaption{\label{bulkbulkboundlot}     Lower     bounds     on    the
compactification scale $M=1/R$ at  the 3$\sigma$ level from  precision
observables and a $\chi^2$ analysis in the SM-bulk model.}
\end{figure}

Figure~\ref{bulkbulkboundlot}  shows      lower      bounds  on    the
compactification   scale    $1/R$ coming    from  different  types  of
observables as functions of  $\sin^2\beta$, where we take into account
only     one   observable   at       a     time.    In       addition,
Fig.~\ref{bulkbulkboundlot} displays the result  obtained by a  global
$\chi^2$ fit. For a model dominated by a brane Higgs field ($\sin\beta
=  1$),   the   most  stringent  bound    on  $1/R$   is  set   by the
forward-backward  asymmetry involving    $b$-quarks,   while   for   a
bulk-Higgs   dominated model (with  $\sin\beta  =  0$)  $1/R$  is most
severely   constrained by the    hadronic  $Z$-boson width.  A  global
$\chi^2$ analysis yields a lower bound on $1/R$ of  about 4~TeV at the
3$\sigma$  CL,  for the  two limiting cases  for which  only one Higgs
field that   either lives in   the bulk or   on the $y=0$  brane has a
non-vanishing VEV.   The lower bound on $1/R$   may decrease to $\sim$
3~TeV, for  a mixed brane-bulk Higgs  scenario  with $\sin^2\beta \sim
0.5$.  This  is so,  because   all  the  observables but  the  various
$Z$-boson widths do not lead  to useful lower  limits on $1/R$ in  the
region of $\sin^2\beta \sim 0.5$.

\subsection{SU(2)$_L$-Brane, U(1)$_Y$-Bulk Model}

Next we shall investigate the model, in which  only the U(1)$_Y$ gauge
boson feels the presence of the extra dimension, whereas the SU$(2)_L$
gauge boson is confined on the $y=0$ brane. In this case, we have
\begin{equation}
\begin{split}
\Delta_Z \, & = \, - \, \hat{s}^2_w \, , \\
\Delta_W \, & = \, 0 \, .
\end{split}
\end{equation}
Obviously, the $W$-boson mass does not change by KK effects.  However,
the modification  of the $Z$-boson  coupling to fermions  becomes more
involved in   this  model.       Specifically, KK   effects     induce
non-factorizable shifts  both in  the   vector and  axial part  of the
$Z\bar{f}f$-coupling,  when  the result is  expressed  in terms of the
$Z$-boson mass-eigenstate. To leading order in $X$, we can account for
these new    non-factorizable   modifications  by  parameterizing  the
$Z\bar{f}f$-coupling   in  terms  of  an   effective   electric charge
$Q_{f (0)}$  and  an  effective  third  component of the  weak isospin
$T_{3 (0)}$:
\begin{equation}
\begin{split}
Q_{f (0)} \ & = \ Q _f\,  \lr{ 1 \, - \, X} \, , \\ 
T_{3f (0)} \ & =\ T_{3f} \, \lr{ 1 \, - \, \hat{s}^2_w \, X} \, ,
\end{split}
\end{equation}
with $Q_f=T_{3f} + Y_f$.   The exact relations between  $Q_{f (0)}$  and
$Q_f$ and between $T_{3f (0)}$ and $T_{3f}$ are given in Appendix~B.
 
Taking the above results into account, we find
\begin{equation}
\Delta_G \ = \ -\, \hat{s}^2_w
\end{equation}
and, thereby,
\begin{equation}
\label{deltathetabranebulk}
\Delta_{\theta} \ = \ \frac{\hat{s}^2_w \, \hat{c}^2_w}{\hat{c}^2_{2w}} \, .
\end{equation}
The simplicity of the above results is a consequence  of the fact that
the charged gauge sector lives on the brane and  hence is not affected
by KK effects.

\begin{table}[t]
\renewcommand{\arraystretch}{1.5}
\begin{center}
\begin{tabular}{c|c} 
\hline \hline Observable & $\Delta^{\mr{HDSM}}_{\mathcal{O}}/X$ \\
\hline \hline $M_W$ & $\frac{1}{2} \lr{\hat{s}^2_w \hat{c}^2_w / \hat{c}_{2w}}$
\\[0.75ex] \hline $\Gamma_Z (\nu \ov{\nu})$ & $- \hat{s}^2_w$ \\
\hline $\Gamma_Z (l^+ l^-)$ & $\hat{s}^2_w \, + \, \Delta_l \, + \,
\delta_l$ \\ \hline $\Gamma_Z (\mr{had})$ & $\hat{s}^2_w \, + \,
\Delta_h \, + \, \delta_h$ \\ \hline $Q_W (\mr{Cs})$ & 
$4 \, Z \, \lr{ Q_W^{\mr{SM}}}^{\! -1} \,
\hat{s}^2_w \, \Delta_{\theta} $ \\[0.75ex] 
\hline $R_{l}$ & $ - \, \Delta_{l} \, + \,
\Delta_h \, - \, \delta_{l} \, + \, \delta_h$ \\ \hline $R_{q}$ & $
\Delta_{q} \, - \, \Delta_h \, + \, \delta_{q} \, - \, \delta_h$ \\
\hline $A_{f}$ & $ \Delta_V- \, \Delta_{f} \, - \delta_f \, + \,
\delta_V \, + \, \delta_A $ \\ [0.75ex] \hline $A_{\mr{FB}}^{(0,f)}$ &
$ \Delta_V- \, \Delta_{f} \, - \delta_f \, + \, \delta_V \, + \,
\delta_A \, + \, f \leftrightarrow e$ \\ [0.75ex] \hline\hline
\end{tabular}
\end{center}
\mycaption{\label{tab2} Predictions for
$\Delta^{\mr{HDSM}}_{\mathcal{O}}/X$ in the SU(2)$_L$-brane,
U(1)$_Y$-bulk model.  See text for the definition of the
delta parameters.}
\end{table}

With  the  help of  the   new  auxiliary   parameters, we exhibit   in
Table~\ref{tab2}        the    tree-level        KK             shifts
$\Delta^{\mr{HDSM}}_{\mathcal{O}}$    to the different     electroweak
observables.   The  parameters $\delta_V$  and  $\delta_A$ give the KK
modifications in the  vector and axial-vector  part of the $Z\bar{f}f$
coupling,  except of  the modifications  which are purely  due to  the
difference between $\theta_w$ and $\hat{\theta}_w$, i.e.
\begin{equation}
\begin{split}
\delta_V \ & = \ \frac{-2 T_{3f} \hat{s}^2_w + 4 Q_f \hat{s}^2_w}{2
T_{3f} - 4 Q_f \hat{s}^2_w} \, ,\\ 
\delta_A \ & = \ - \hat{s}^2_w \, .
\end{split}
\end{equation}
The  parameter $\delta_f$ quantifies the  KK  shift in  the sum of the
squared vector  and axial vector couplings  of a given fermion  $f$ to
the   $Z$ boson  in  this  SU(2)$_L$-brane,  U(1)$_Y$-bulk model.  The
parameter $\delta_f$ is given by
\begin{equation}
\delta_f \ = \ \frac{\lr{- 16 T^2_{3f} + 16 T_{3f} Q_f} \hat{s}^2_w
                                            + \lr{ 16 T_{3f} Q_f - 32
                                            Q_f^2} \hat{s}^4_w} {(2
                                            T_{3f} - 4 Q_f
                                            \hat{s}^2_w)^2 + (2
                                            T_{3f})^2} \, .
\end{equation}
In analogy with $\Delta_h$, we finally define $(q=u,d,c,s,b)$ 
\begin{equation}
\delta_h \ = \ \frac{ \sum_{q} \left[ \lr{- 16 T^2_{3f} + 
16 T_{3f} Q_f} \hat{s}^2_w + \lr{ 16 T_{3f} Q_f - 32 Q_f^2} 
\hat{s}^4_w \right]}{ \sum_{q} \left[ (2 T_{3f} - 4 Q_f 
\hat{s}^2_w)^2 + (2 T_{3f})^2 \right]} \, .
\end{equation}
Moreover, the  parameters  $\Delta_V$, $\Delta_f$  and  $\Delta_h$ are
defined in  \equ{deltadefinitions}  with  $\Delta_{\theta}$  given  by
\equ{deltathetabranebulk}.

\begin{table}[t]
\renewcommand{\arraystretch}{1.5}
\begin{center}
\begin{tabular}{c|c|c} 
\hline\hline 
Observable & U(1)$_Y$ in bulk & SU(2)$_L$ in bulk \\ \hline \hline   
$M_W$ & 1.2 & 1.2\\ \hline
$\Gamma_Z (\mr{had})$ & 0.8 & 2.3 \\ \hline 
$Q_W (\mr{Cs})$ & 0.4 & 0.8 \\ \hline
$A_{\mr{FB}}^{(0,b)}$ & 4.4 & 2.4 \\ \hline
$A_{\tau}$ & 2.5 & 1.4 \\ \hline
$R_{\tau}$ & 1.0 & 0.5 \\ \hline
\parbox{3cm}{\vspace{0.2cm} \begin{center} global analysis
\end{center} \vspace{0.2cm}} & 3.5 & 2.6 \\ 
\hline\hline
\end{tabular}
\end{center}
\mycaption{\label{branebulkbounds}  Lower   bounds   (in  TeV)  on the
compactification scale $1/R$   at  the 3$\sigma$ CL  in  models where
either only the U(1)$_Y$ or only  the SU(2)$_L$ gauge boson propagates
in the higher-dimensional space.}
\end{table}

Following  the procedure outlined in the  previous section, we can now
evaluate the lower bounds on the compactification scale $M=1/R$ in the
SU(2)$_L$-brane, U(1)$_Y$-bulk model.  In Table~\ref{branebulkbounds},
we display the  lower limits on  $1/R$ for each observable separately,
together with that  found by a global  analysis.  The most restrictive
bound is obtained by  the $b$-quark forward-backward asymmetry, giving
rise to a lower limit on $1/R$ of $\sim 4.4$~TeV  at the $3\sigma$ CL.
Finally,    our   global-fit analysis   leads   to  the  slightly less
restrictive lower bound: $1/R \stackrel{>}{{}_\sim}$~3.5~TeV.

\subsection{SU(2)$_L$-Bulk, U(1)$_Y$-Brane Model}

Let us finally consider the complementary  scenario, in which only the
SU(2)$_L$ gauge boson propagates in the  higher-dimensional space.  In
this case, the KK-mass  shifts  for the $Z$   and $W^\pm$  bosons  are
computed to be
\begin{equation}
\Delta_Z \ = \ \Delta_W\ =\ - \, \hat{c}^2_w \, .
\end{equation}
By analogy,  the KK effects  on  the $Z\bar{f}f$-coupling can  also be
taken into account by introducing  an effective third component of the
weak isospin:
\begin{equation}
T_{3f (0)} \ \ =\ T_{3f} \, \lr{ 1 \, -
\, \hat{c}^2_w \, X} \, .
\end{equation}
Unlike  in   the   model discussed    in  the  previous   section, the
electric-charge term in the $Z\bar{f}f$-coupling remains unaffected by
KK effects, i.e.\ $Q_{f (0)} \ =\ Q _f$. Thus, from the muon decay, we
calculate
\begin{equation}
\Delta_G \ = \ - \, \hat{c}^2_w,
\end{equation}
which leads to
\begin{equation}
  \label{Dthe}
\Delta_{\theta} \ = \ \frac{\hat{c}^4_w}{\hat{c}^2_{2w}} \, .
\end{equation}

\begin{table}[t]
\renewcommand{\arraystretch}{1.5}
\begin{center}
\begin{tabular}{c|c} 
\hline\hline 
Observable & $\Delta^{\mr{HDSM}}_{\mathcal{O}}/X$ \\ \hline \hline
$M_W$ & $\frac{1}{2} \lr{\hat{s}^2_w \hat{c}^2_w / \hat{c}_{2w}}$ \\[0.75ex] \hline
$\Gamma_Z (\nu \ov{\nu})$ & $- \, \hat{c}^2_w$ \\ \hline $\Gamma_Z
(l^+ l^-)$ & $\hat{c}^2_w \, + \, \Delta_l \, + \, \delta_l$ \\ \hline
$\Gamma_Z (\mr{had})$ & $\hat{c}^2_w \, + \, \Delta_h \, + \,
\delta_h$ \\ \hline $Q_W (\mr{Cs})$ & 
$4 \, Z \, \lr{ Q_W^{\mr{SM}}}^{\! -1} \,
\hat{s}^2_w \, \Delta_{\theta} $ \\[0.75ex] \hline $R_{l}$ & 
$ - \, \Delta_{l} \, + \,
\Delta_h \, - \, \delta_{l} \, + \, \delta_h$ \\ \hline $R_{q}$ & $
\Delta_{q} \, - \, \Delta_h \, + \, \delta_{q} \, - \, \delta_h$ \\
\hline $A_{f}$ & $ \Delta_V- \, \Delta_{f} \, - \delta_f \, + \,
\delta_V \, + \, \delta_A $ \\ [0.75ex] \hline $A_{\mr{FB}}^{(0,f)}$ &
$ \Delta_V- \, \Delta_{f} \, - \delta_f \, + \, \delta_V \, + \,
\delta_A \, + \, f \leftrightarrow e$ \\ [0.75ex] 
\hline\hline
\end{tabular}
\end{center}
\mycaption{\label{tab3} Predictions for
$\Delta^{\mr{HDSM}}_{\mathcal{O}}/X$ in the SU(2)$_L$-bulk,
U(1)$_Y$-brane model.  See text for the definition of the auxiliary
parameters.}
\end{table}

As in the   previous section, we  introduce the   auxiliary parameters
$\Delta_V$, $\Delta_f$, $\Delta_h$, $\delta_V$, $\delta_A$, $\delta_f$
and $\delta_h$,  which  enables us to  cast the  tree-level KK  shifts
$\Delta^{\mr{HDSM}}_{\mathcal{O}}$ to  the  electroweak observables in
Table~\ref{tab3}.  The meaning  of these auxiliary parameters  are the
same  as  in Sections~5.1  and     5.2.  In  particular,   $\Delta_V$,
$\Delta_f$,   $\Delta_h$  are given  by   \equ{deltadefinitions}, with
$\Delta_\theta$        in~\equ{Dthe},   while $\delta_V$,  $\delta_A$,
$\delta_f$   and   $\delta_h$   are   respectively    found    to   be 
($q=u,d,c,s,b$)
\begin{equation}
\begin{split}
\delta_V & = -\, \frac{2 T_{3f} \hat{c}^2_w }{2 T_{3f} - 4 Q_f
\hat{s}^2_w} \, , \\[1ex] 
\delta_A & = -\, \hat{c}^2_w \, , \\[1ex]
\delta_f & = \frac{\lr{- 16 T^2_{3f} + 16 T_{3f} Q_f \hat{s}^2_w
} \hat{c}^2_w } {(2 T_{3f} - 4 Q_f \hat{s}^2_w)^2 + (2 T_{3f})^2}
\, , \\[1ex]
\delta_h  & = \frac{\sum_{q} \lr{- 16 T^2_{3f} + 
16 T_{3f} Q_f \hat{s}^2_w} \hat{c}^2_w} {\sum_{q} \left[ 
(2 T_{3f} - 4 Q_f \hat{s}^2_w)^2 + (2 T_{3f})^2 \right] } \, .
\end{split}
\end{equation}

In Table~\ref{branebulkbounds}, we  also  present the lower  bounds on
$1/R$ for the different type of observables. In the present model, the
$b$-quark forward-backward  asymmetry offers the  most stringent lower
bound   on      the    compactification    scale    as     well:  $1/R
\stackrel{>}{{}_\sim}   2.4$~TeV    at   the     3$\sigma$ CL.    Most
interestingly, we observe that this lower  bound on $1/R$ is much more
relaxed  than   the  one  found  in  the   previous  models.  The same
observation  applies  to  our  global  fit  as  well,  i.e.~a $\chi^2$
analysis constrains the compactification   scale $M=1/R$ to  be higher
than about~2.6~TeV at the 3$\sigma$ CL.

In Table~\ref{globalbounds}, we summarize  the  lower bounds on  $1/R$
obtained  by  our   global  fits  in   the  minimal higher-dimensional
extensions of the  SM under  discussion.  We  find that  the  $\chi^2$
values increase rapidly  as the compactification scale decreases, such
that  the  lower  bounds  on $1/R$   at higher  confidence levels  are
relatively stable.  Thus, from  Table~\ref{globalbounds}, we see again
that the lower  bound on the compactification  scale is the smallest in
the SU(2)$_L$-bulk, U(1)$_Y$-brane model.

\begin{table}[t]
\renewcommand{\arraystretch}{1.5}
\begin{center}
\begin{tabular}{c|c|c|c} 
\hline\hline 
model & 2$\sigma$ & 3$\sigma$ & 5$\sigma$ \\ \hline \hline   
\parbox{5cm}{\vspace{0.2cm} \begin{center} SU(2)$_L$-brane, 
U(1)$_Y$-bulk \end{center} \vspace{0.2cm} } & \hspace{0.2cm} 4.3
\hspace{0.2cm} & \hspace{0.2cm} 3.5 \hspace{0.2cm} & \hspace{0.2cm}
2.7 \hspace{0.2cm} \\ \hline
\parbox{5cm}{\vspace{0.2cm} \begin{center} SU(2)$_L$-bulk, 
U(1)$_Y$-brane \end{center} \vspace{0.2cm} } & 3.0 & 2.6 & 2.1 \\ \hline
\parbox{5cm}{\vspace{0.2cm} \begin{center} SU(2)$_L$-bulk, 
U(1)$_Y$-bulk (brane Higgs) \end{center} \vspace{0.2cm} } & 4.7 
& 4.0 & 3.1 \\ \hline
\parbox{5cm}{\vspace{0.2cm} \begin{center} SU(2)$_L$-bulk, 
U(1)$_Y$-bulk (bulk Higgs) \end{center} \vspace{0.2cm} } & 4.6 
& 3.8 & 3.0 \\ \hline\hline
\end{tabular}
\end{center}
\mycaption{\label{globalbounds} Lower bounds (in TeV) on the
compactification scale $1/R$ at 2$\sigma$, 3$\sigma$ and 5$\sigma$
CLs.}
\end{table}

\setcounter{equation}{0}
\section{Conclusions}

\indent

We  have studied  new   possible 5-dimensional  extensions of   the SM
compactified  on an $S^1/Z_2$ orbifold,   in  which the SU(2)$_L$  and
U(1)$_Y$ gauge  fields and Higgs bosons may  or may not all experience
the presence of the fifth dimension.  Moreover,  the fermions in these
models are considered to be confined to the  one of the two boundaries
of the $S^1/Z_2$  orbifold.    We  have  paid special    attention  to
consistently quantize the higher-dimensional models in the generalized
$R_\xi$  gauges.  Specifically, we have   been  able to identify   the
appropriate higher-dimensional gauge-fixing conditions which should be
imposed on  the theories so as to  yield the known $R_\xi$ gauge after
the   fifth   dimension  has   been integrated   out.    Based  on the
so-quantized   effective  Lagrangians,    we  have derived    analytic
expressions for the KK-mass spectrum of the gauge bosons and for their
interactions to the fermionic matter.

The aforementioned analytic  expressions have proven very essential to
obtain  accurate predictions  for  low-energy as  well  as high-energy
electroweak observables measured at  LEP  and SLC.  In  particular, we
have  performed  an    extensive   global-fit   analysis   of   recent
high-precision  electroweak data   to  three different   5-dimensional
extensions of  the  SM:  (i)~the  SU(2)$_L\otimes$U(1)$_Y$-bulk model,
where all SM  gauge bosons are  bulk fields; (ii) the SU(2)$_L$-brane,
U(1)$_Y$-bulk model, where only the  $W^\pm$ bosons are restricted  to
the brane, and  (iii)~the SU(2)$_L$-bulk, U(1)$_Y$-brane  model, where
only  the  U(1)$_Y$  gauge field  is  confined  to  the  brane.  After
carrying out a $\chi^2$-test, we obtain different sensitivities to the
compactification radius  $R$   for the above  three models.    For the
often-discussed first model, we  find the 2$\sigma$ (3$\sigma$)  lower
bounds  on     $1/R$:     $1/R\stackrel{>}{{}_\sim}  4.6$~(3.6)    and
4.7~(4.0)~TeV, for a Higgs boson living in the bulk  and on the brane,
respectively.   For the second  and   third models, the  corresponding
2$\sigma$ (3$\sigma$)  lower  limits are 4.3~(3.5)~and  3.0~(2.6)~TeV.
Consequently, we observe that the bounds  on $1/R$ may be reduced even
up to 1~TeV, if the $W^\pm$ bosons are  the only fields that propagate
in the bulk.

The analysis presented here involves a number of assumptions which are
inherent   in    any   non-stringy   field-theoretic     treatment  of
higher-dimensional theories.  Although  the  results obtained  in  the
higher-dimensional models with one compact dimension are convergent at
the    tree level, they  become   divergent  if  more  than one  extra
dimensions are considered. Also, the analytic results are ultra-violet
(UV) divergent at the   quantum  level, since the   higher-dimensional
theories are not renormalizable.  Within a string-theoretic framework,
the above UV divergences are expected to  be regularized by the string
mass scale $M_s$.  Therefore, from an  effective field-theory point of
view,  the phenomenological predictions will depend  to some extend on
the UV cut-off procedure~\cite{KMZ} related to the string scale $M_s$.
Nevertheless, assuming validity of perturbation theory, we expect that
quantum  corrections due to extra  dimensions will not exceed the 10\%
level of the tree-level effects we have been  studying here.  Finally,
we have ignored possible model-dependent winding-number contributions,
which become relevant when the compactification  scale $1/R$ and $M_s$
turn out to be of comparable size~\cite{ABL}.

The lower limits on the compactification scale  derived by the present
global  analysis  indicate that resonant  production of   the first KK
state may only be accessed at the LHC, at  which heavy KK masses up to
6--7~TeV~\cite{AB,RW} might  be  explored.    In particular,   if  the
$W^\pm$ bosons propagate in the bulk with a compactification radius $R
\sim  3$~TeV$^{-1}$, one may  still be able  to probe resonant effects
originating from  the second KK  state, and so differentiate the model
from other 4-dimensional new-physics scenaria.

\subsection*{Acknowledgements}
This  work was supported  by the  Bundesministerium f\"ur  Bildung and
Forschung (BMBF,  Bonn, Germany) under the  contract numbers 05HT9WWA9
and 05HT1WWA2.

\subsection*{Note Added}
Shortly  after completion of our  paper, we became  aware of \cite{MN}
and \cite{CL}. The   focus of these   papers is the SM-bulk model,  in
which KK effects on high-energy scattering processes at LEP2 and other
colliders were  analyzed.  In   addition  to being  complementary   by
concentrating   on   high-precision  electroweak  observables, we have
investigated  new minimal   higher-dimensional extensions   of the SM,
where  the SU$(2)_L$ and U(1)$_Y$ gauge  bosons may not both propagate
in the higher-dimensional space. In particular, we find that the lower
limits   on $1/R$ may  be   substantially  relaxed  in  one  of  these
scenarios.  Finally, we address the issue of a consistent quantization
of the higher-dimensional  field  theory  in the  generalized  $R_\xi$
gauge.

\newpage

\def\theequation{\Alph{section}.\arabic{equation}}
\begin{appendix}

\setcounter{equation}{0}
\section{Goldstone Modes in the Abelian 2-Higgs Model}
\label{Goldstonemodes}

In  this appendix, we  wish to show  that the KK Goldstone modes given
in~\equ{wouldbegoldstonemodesgeneralabelianhiggs}  have the properties
of   true Goldstone particles  as  these   are known from  spontaneous
symmetry   breaking  models.    The higher-dimensional    gauge-fixing
Lagrangian in~\equ{gaugefixingfunctiongenabelianmodel} induces at each
KK level $n$ the gauge fixing terms
\begin{equation}
\label{rxigaugefixingtermsabmodelgeneralcase}
\la^{(n)}_{\mr{GF}} \ = \ - \frac{1}{2 \xi} \, \bigg[\,\partial_{\mu}
                             A^{\mu}_{(n)} \, - \, \xi \,
                             \bigg(\sqrt{\frac{n^2}{R^2} + e^2 v^2 \cos^2
                             \beta } \, \, G_{(n)} \, + \, \sqrt{2}^{1-
                             \delta_{n,0}} e v \sin\beta\, 
                             \chi_2 \,  \bigg)\, \bigg]^2 \, ,
\end{equation}
where the factor of  $\sqrt{2}$ stems from the $\delta$-function (cf.\
\equ{deltafunctionrepres}).  In the  Abelian 2-Higgs model, the fields
$G_{(n)}$         are         defined         analogously         with
\equ{physunphysfieldshiggsinbulk} as
\begin{equation}
G_{(n)}\ = \ 
\bigg( \frac{n^2}{R^2}\: +\: e^2 v^2 \cos^2 \beta\bigg)^{-1/2}
\bigg(\,\frac{n}{R}\, A_{(n) 5}\: + \: ev \cos \beta\,
\chi_{1(n)}\,\bigg)\, .
\end{equation}
Thus, the $\xi$-dependent  mass  terms  of  the  scalar modes  in  the
$\chi_2 G_{(n)}$-basis are given by
\begin{equation}
\label{unphysicalmasstermsgeneralhiggsmodel}
\la^\xi_{\mr{mass}} (x)\ =\ 
- \frac{\xi}{2} \left( \chi_2 , \, G_{(0)} , \, G_{(1)} , \, \ldots \right)
M_{\xi}^2
\left(
\begin{array}{c}
\chi_2 \\ 
G_{(0)} \\
G_{(1)} \\
\vdots
\end{array}
\right) \, ,
\end{equation}
with
\begin{equation}
M_{\mr{\xi}}^2\ =\ 
\left(
\begin{array}{cccc}
e^2 v^2 \, \lr{1 + \sum^{\infty}_{n=1} \, 2} \, \ssb & e^2 v^2 \sib
\cob & \sqrt{2} \, e v \, c_1 \, \sib & \cdots \\ e^2 v^2 \sib \cob &
e^2 v^2 \csb & 0 & \cdots \\ \sqrt{2} \, e v \, c_1 \, \sib & 0 &
\lr{1/R}^2 + e^2 v^2 \csb & \cdots \\ \vdots & \vdots & \vdots &
\ddots
\end{array}
\right)
\end{equation}
and $c_n =  \sqrt{\lr{n/R}^2 + \lr{e v  \cos  \beta}^2}$. The infinite
sum in  the  upper left entry   of $M_{\xi}^2$  is due  to $\delta(0)$
according to \equ{deltazerorepres}.  We expect that only the Goldstone
modes  of the theory  acquire  gauge-dependent masses coming from  the
gauge-fixing  terms.   Computing   the   characteristic  polynomial of
$M^2_\xi$, we find
\begin{equation}
\label{characteristicpolyofmphysandmunphys}
\det ( M_{\xi}^2 - \lambda {\rm I}) \ = \ - \, \lambda \, \det (
M_A^2 - \lambda {\rm I}) \, ,
\end{equation}
where $M^2_A$ is the gauge-boson mass matrix given in~\equ{M2A}.  As a
consequence, we may assign a Goldstone mass eigenstate $\hat{G}_{(n)}$
with  mass  $\sqrt{\xi} m_{A (n)}$  for  each  KK gauge eigenmode with
mass  $m_{A (n)}$.   This  constitutes  a necessary condition in order
to obtain  gauge-invariant S-matrix elements within  the $R_\xi$ class
of   gauges.    {}From~\equ{characteristicpolyofmphysandmunphys},   we
observe the  existence of an  additional degree of freedom  which does
not  acquire a  $\xi$-dependent mass  with no  correspondence to  a KK
gauge  mode.   This  additional  CP-odd scalar  field  will  generally
receive  a gauge-independent  mass that  will entirely  depend  on the
parameters of the  Higgs potential. Additionally, it may  mix with the
other physical CP-odd states  to form mass eigenstates (see discussion
below).

On  the other hand,  in a  consistent theory,  the KK  Goldstone modes
should not  acquire any gauge-independent  mass term apart  from their
$\xi$-dependent  mass mentioned  above.  In  addition to  the  KK mass
terms, the physical  mass matrix of the KK  scalar modes is determined
by the Higgs kinetic terms in~\equ{lageneralabelianhiggsmodel} and the
Higgs potential~\equ{generalhiggspotential}.   Since the CP-even Higgs
modes do not mix with $A_{(n)5}$ in the CP-conserving case, the scalar
mass matrix  is block  diagonal and we  can concentrate on  the CP-odd
mass  matrix  $M_{\mr{CP-odd}}^2$,  as  it  appears  in  the  original
Lagrangian
\begin{equation}
\la^{\mr{CP-odd}}_{\mr{mass}} (x,y)\ =\ 
- \frac{1}{2} \left( \chi_1 , \, \chi_2 \right)
M_{\mr{CP-odd}}^2
\left(
\begin{array}{c}
\chi_1 \\ 
\chi_2 \\
\end{array}
\right) \, ,
\end{equation}
before  integrating  out the  $y$ dimension.  After  a  straightforward 
computation from \equ{generalhiggspotential},  this CP-odd mass matrix 
may be cast into the form
\begin{equation}
\label{fivedimcpoddmassmatrix}
M_{\mr{CP-odd}}^2 \ = \ \delta (y)\,
\left(
\begin{array}{cc}
m^2_{\chi 11} & m^2_{\chi 12}\\
m^2_{\chi 12} & m^2_{\chi 22}\\
\end{array}
\right)\ \, 
\end{equation}
where
\begin{equation}
\label{mchi11}
m^2_{\chi 11} \ =\ - \, \tan \beta \, m_{12}^2\: +\: 2 v^2 \ssb \,
\lambda_5\:  +\: \frac{1}{2} v^2 \sib \cob \, \lambda_6\: 
                   + \: \frac{1}{2} v^2 \ssb \, \tan \beta \,
\lambda_7\, .
\end{equation}
The other entries of the CP-odd mass matrix $M_{\mr{CP-odd}}^2$ can be
related to $m^2_{\chi 11}$ via
\begin{equation}
  \label{mchi12}
m^2_{\chi 22} \ = \ m^2_{\chi 11}/ \tan^2 \beta \quad \mr{and} \quad 
m^2_{\chi 12} \ = \ m^2_{\chi 21}\ =\ - \, m^2_{\chi 11}/ \tan \beta \, .
\end{equation}
In         deriving~\equ{fivedimcpoddmassmatrix},         \equ{mchi11}
and~\equ{mchi12}, we  have made use of the  minimization conditions on
the Higgs  potential, i.e.~$\big<  \partial V/\partial \Phi_i  \big> =
0$, with  $i= 1,2$. In particular,  the latter enabled us  to cast the
CP-odd     mass     matrix     into     the     simple     form     of
\equ{fivedimcpoddmassmatrix}, where all entries are proportional to an
overall $\delta$-function.   Note that the absence of  bulk mass terms
originating from the Higgs potential is a characteristic of the CP-odd
scalar sector of the model under consideration.

After       integrating        out       the       $y$       dimension
in~\equ{fivedimcpoddmassmatrix}, we  obtain the effective  mass matrix
for  all the  CP-odd KK  modes  $\chi_{1 (n)}$,  $\chi_2$ and  $A_{(n)
5}$.  From this  effective CP-odd  mass matrix  including the  KK mass
terms, it  is straightforward, although somehow tedious,  to show that
the                         would-be                         Goldstone
modes~\equ{wouldbegoldstonemodesgeneralabelianhiggs}  do  not  receive
indeed any  gauge-independent mass  from the Higgs  potential, whereas
all physical  CP-odd mass eigenstates should acquire  high enough mass
eigenvalues to avoid conflict with experimental data.

\setcounter{equation}{0}
\section{Masses, Couplings and Feynman rules}
\label{Lagrangianmassescouplings}

Here, we shall  present exact analytic results for  the masses and the
couplings  of  the  KK  gauge    modes to   fermions in  the   minimal
5-dimensional extensions of the SM discussed in Section~4.

\begin{figure}[t]
\begin{center}
\begin{picture}(450,270)(0,50)
\SetWidth{1.}
\thicklines

\scalarpropagator[$\hat{G}^{\pm}_{(n)}$ propagator:]{160}{50}{\mu}{(n)}{\nu}{=
\frac{i}{k^2 - \, \xi \, m_{W (n)}^2}}

\scalarpropagator[$\hat{G}^0_{(n)}$ propagator:]{160}{100}{\mu}{(n)}{\nu}{=
\frac{i}{k^2 - \, \xi \, m_{Z (n)}^2}}

\scalarpropagator[$A_{(n)5}$ propagator:]{160}{150}{\mu}{(n)}{\nu}{=
\frac{i}{k^2 - \, \xi \, m^2_{\gamma (n)} }}

\gaugepropagator[$\hat{Z}_{(n)}$-boson propagator:]{160}{200}{\mu}{(n)}{\nu}{=
\frac{i}{k^2 - m_{Z (n)}^2} \, \left( - g^{\mu \nu} \, + \, \,
\frac{ \left( 1 - \xi \right) k^{\mu} \, k^{\nu}}{ k^2 - \, \xi \,
m_{Z (n)}^2} \, \right)}

\gaugepropagator[$\hat{W}^\pm_{(n)}$-boson propagator:]{160}{250}{
\mu}{(n)}{\nu}{=
\frac{i}{k^2 - m_{W (n)}^2} \, \left( - g^{\mu \nu} \, + \, \,
\frac{ \left( 1 - \xi \right) k^{\mu} \, k^{\nu}}{ k^2 - \, \xi \,
m_{W (n)}^2} \, \right)}

\gaugepropagator[$\gamma_{(n)}$ propagator:]{160}{300}{\mu}{(n)}{\nu}{=
\frac{i}{k^2 - m^2_{\gamma (n)}} \, \left( -
g^{\mu \nu} \, + \, \, \frac{ \left( 1 - \xi \right) k^{\mu} \,
k^{\nu}}{ k^2 - \, \xi \, m^2_{\gamma (n)}} \,
\right)}

\end{picture}
\end{center}
\mycaption{\label{propagatorsin5DGSW} KK gauge-   and  Goldstone-boson
propagators   in  the  5-dimensional   extensions  of  the SM  in  the
generalized $R_\xi$-gauge. }
\end{figure}

To  start    with, we  display   in Fig.~\ref{propagatorsin5DGSW}  the
propagators for the KK gauge and Goldstone modes in the $R_\xi$ gauge.
In addition, the masses  of the KK gauge bosons  may be determined as
follows:

\noindent
{\bf SU(2)$_L \otimes $U(1)$_Y$-Bulk Model:}
\begin{eqnarray}
m_{\gamma (n)} &=& \frac{n}{R}\, ,\\
\sqrt{m^2_{W (n)} - m^2_W \cos^2 \beta} &=& 
\pi m^2_W \sin^2 \! \beta \, 
R \, \cot \Big(\,\pi R \, 
\sqrt{m^2_{W (n)} - m^2_W \cos^2 \beta}\,\Big) \,,\\
\sqrt{m^2_{Z (n)} - m^2_Z \cos^2 \beta} &=& \pi m^2_Z \sin^2 \!
               \beta \, R \, \cot \Big(\,\pi R \, \sqrt{m^2_{Z
               (n)} - m^2_Z \cos^2 \beta}\,\Big) \, ,
\end{eqnarray}
where $n=0,1,2,\ldots$, $m_W = g v / 2 $ and $m_Z =
\sqrt{g^2+g^{\prime 2}}\, v/2$.

\noindent
{\bf SU(2)$_L$-Brane, U(1)$_Y$-Bulk Model:}
\begin{eqnarray}
m_{Z (n)} & =& \pi m^2_Z \, \sin^2 \theta_w \, R \, \cot
\big(\,\pi R \, m_{Z (n)}\,\big) \, + \,  
\frac{m^2_Z}{m_{Z (n)}} \, \cos^2 \theta_w\ . 
\end{eqnarray}
Note that there are no KK excitations for the photon  and $W$ boson in
this model.

\noindent
{\bf SU(2)$_L$-Bulk, U(1)$_Y$-Brane Model:}
\begin{eqnarray}
m_{W (n)} &=& \pi m^2_W 
                \, R \, \cot \Big(\,\pi R \, m_{W (n)}\,\Big) 
		\, , \\  
m_{Z (n)} &=&   \pi m^2_Z \, \cos^2 \theta_w \, R \, \cot \big(\,\pi R
                \, m_{Z (n)}\,\big) \, + \, 
		\frac{m^2_Z}{m_{Z (n)}} \sin^2 \theta_w \ . 
\end{eqnarray}
There are no KK excitations for the photon field in this model.

In the following, we will give the  exact analytic expressions for the
couplings  of KK  gauge bosons  to fermions.   To  this end, we  first
define the following generic interaction Lagrangian:
\begin{equation}
\label{finalgenericlagrangian}
\la_{\mr{int}}\, = \, \sum_{n} \, g_{W (n)} \lr{
\hat{W}^{+}_{(n) \mu} \, J^{+ \mu}_{W} \, + \, \hat{W}^{-}_{(n) \mu}
\, J^{- \mu}_{W} } \, + \, \sum_{n} \, g_{Z (n)} \,
\hat{Z}_{(n)\mu} \, J^{\mu}_{Z} \, + \, \sum_{n} \, e_{(n)} \,
\hat{A}_{(n)\mu} \, J^{\mu}_{\mr{EM}} \, ,
\end{equation}
with
\begin{equation}
\label{currentsapp}
\begin{split}
J^{+ \mu}_{W} \ & = \ \frac{1}{2 \sqrt{2}} \, \lreckig{ \ov{\nu}_{i}
            \, \gamma^{\mu} \, \lr{1 - \gamma^{5}} \, e_i \, + \,
 \ov{u}_i \, \gamma^{\mu} \,  \lr{1 -\gamma^{5}} \, d_j \, \, V_{ij}}\,, \\
J^{\mu}_{Z} \ &  = \ \frac{1}{4\cos \theta_w} \,
      \ov{f} \, \gamma^{\mu} \, \lreckig{ \lr{2 \, T_{3f (n)}
      - 4 \, Q_{f (n)} \sin^2 \theta_w} \, - \, 2 \, T_{3f (n)}
      \, \gamma^{5}} \, f  \,, \\
J^{\mu}_{EM} \ &  = \ \ov{f} \, Q_f \, \gamma^{\mu} \, f
\end{split}
\end{equation}
and $\nu_{i} = (\nu_e, \, \nu_{\mu}, \,  \nu_{\tau})$, $e_{i} = (e, \,
\mu, \, \tau)$, $u_{i} =  (u, \, c, \,  t)$ and $d_{i}  = (d, \, s, \,
b)$.  In addition, $f$ denotes all the 12 SM  fermions.  After a basis
transformation from  the weak to the  mass eigenstates, we  obtain the
following effective  gauge and quantum couplings  related to the three
different higher-dimensional models ($n= 0,1,2,\ldots$):

\noindent
{\bf SU(2)$_L \otimes $U(1)$_Y$-Bulk Model:}
\begin{equation}
\begin{split}
& \quad \quad \quad \quad \quad e_{(0)} \ = \ e \, , \quad
\quad \quad \quad \quad \quad e_{(n \ge 1)} \ = \ \sqrt{2} \,
e \, , \\ 
g_{Z (n)} \ & = \ \sqrt{2} \, g \, \bigg(\,1 \ + \
\frac{m_Z^2 \ssb}{m^2_{Z (n)} - m_Z^2 \csb} \ + \
\frac{\pi^2\,m_Z^4 \sin^4 \beta}{M^2\,(m^2_{Z (n)} - m_Z^2 \csb)}
\,\bigg)^{-1/2},\\ 
g_{W (n)} \ & = \ \sqrt{2} \, g \, \bigg(\,1
\ + \ \frac{m_W^2 \ssb}{m^2_{W (n)} - m_W^2 \csb} \ + \
\frac{\pi^2\,m_W^4 \sin^4 \beta}{M^2\,(m^2_{W (n)} - m_W^2\csb)} 
\,\bigg)^{- 1/2},\\[1ex] & \quad \quad \quad \quad \quad
T_{3f (n)} \ = T_{3f} \, \, , \quad \quad \quad \quad \quad
Q_{f (n)} \ =\ Q_{f}\, ,
\end{split}
\end{equation}
with $M=1/R$.

\noindent
{\bf SU(2)$_L$-Brane, U(1)$_Y$-Bulk Model:}
\begin{equation}
\begin{split}
& \quad \quad \quad \quad \quad \quad \quad \quad \quad \quad \quad
\quad g_{Z (n)} \ = \ g \, , \\ T_{3f (n)}\ & =\
\frac{T_{3f}}{c_w} \, \frac{m^2_{Z (n)}}{m^2_{Z}} \,
\bigg[\,\frac{1}{s^2_w} \bigg(\frac{1}{2} - \frac{m^2_{Z
(n)}}{m^2_{Z}}\bigg) + \frac{s^2_w}{2c^2_w} \bigg(\pi^2 \frac{m^2_{Z
(n)}}{M^2} +\frac{m^2_{Z (n)}}{m^2_{Z} s^2_w} 
+ \frac{m^4_{Z (n)}}{m^4_{Z} s^4_w}\bigg) \bigg]^{-1/2}, \\
Q_{f (n)}\ & =\, \frac{Q_{f}}{c_w} \bigg( \frac{m^2_{Z
(n)}}{m_Z^2 s^2_w} - \frac{c^2_w}{s^2_w}\bigg)\bigg[\,\frac{1}{s^2_w}
\bigg( \frac{1}{2} - \frac{m^2_{Z (n)}}{m^2_{Z}}\bigg) +
\frac{s^2_w}{2c^2_w} \bigg(  \pi^2 \frac{m^2_{Z (n)}}{M^2} 
+ \frac{m^2_{Z(n)}}{m^2_{Z} s^2_w}  +
\frac{m^4_{Z (n)}}{m^4_{Z} s^4_w} \bigg)\, \bigg]^{-1/2}.
\end{split}
\end{equation}

\noindent
{\bf SU(2)$_L$-Bulk, U(1)$_Y$-Brane Model:}
\begin{equation}
\begin{split}
& \quad \quad g_{Z (n)} \ = \ g\,, \quad
\quad g_{W (n)} \ = \ \sqrt{2} \, g \, \bigg(  1 \, + \,
\frac{m_W^2}{m^2_{W (n)}} \, + \, 
\frac{\pi^2 m_W^4}{M^2 m^2_{W (n)}} \bigg)^{-1/2},\\[1ex] 
T_{3f (n)}\ & =\
\frac{T_{3f}}{s_w} \, \frac{m^2_{Z (n)}}{m^2_{Z}} \,
\bigg[\,\frac{1}{c^2_w} \bigg(\frac{1}{2} - \frac{m^2_{Z
(n)}}{m^2_{Z}}\bigg) + \frac{c^2_w}{2s^2_w} \bigg( \pi^2
\frac{m^2_{Z (n)}}{M^2} + \frac{m^2_{Z (n)}}{m^2_{Z}c^2_w} 
+ \frac{m^4_{Z (n)}}{m^4_{Z}c^4_w}\bigg)\,
\bigg]^{-1/2}, \\
Q_{f (n)}\ & =\ \frac{Q_{f}}{s_w} 
\bigg[\,\frac{1}{c^2_w} \bigg(\frac{1}{2} - \frac{m^2_{Z
(n)}}{m^2_{Z}}\bigg) + \frac{c^2_w}{2s^2_w} \bigg( \pi^2
\frac{m^2_{Z (n)}}{M^2} + \frac{m^2_{Z (n)}}{m^2_{Z}c^2_w}
+ \frac{m^4_{Z (n)}}{m^4_{Z}c^4_w} \bigg)\,
\bigg]^{-1/2}.
\end{split}
\end{equation}

\begin{figure}[t]
\begin{center}
\begin{picture}(450,275)(30,45)
\SetWidth{1.}
\thicklines

\threevertex[]{100}{75}{{vector}{$$}{
$\hat{Z}^{\mu}_{(n)}$}}{{fermionout}{}{$f$}}{{fermionin}{}{$f$}}
{\quad \quad = \, \frac{i }{4 \cos \theta_w} \, g_{Z (n)} \,
\gamma^{\mu} \, \lreckig{ \lr{2 T_{3 f (n)} - 4 Q_{f (n)}
\sin^2 \theta_w} - 2 T_{3 f (n)} \gamma^5}}

\threevertex[]{100}{150}{{vector}{$$}{$\hat{W}^{-
\mu}_{(n)}$}}{{fermionout}{}{$\! d_i$}}{{fermionin}{}{$\! u_j$}}
{\quad \quad = \, \frac{i }{2 \sqrt{2}} \, g_{W (n)} \,
\gamma^{\mu} \, \lr{ 1 - \gamma^5} V^*_{ji}}

\threevertex[]{100}{225}{{vector}{$$}{$\hat{W}^{-
\mu}_{(n)}$}}{{fermionout}{}{$e_i$}}{{fermionin}{}{$\nu_{i}$}} {\quad
\quad = \, \frac{i }{2 \sqrt{2}} \, g_{W (n)} \, \gamma^{\mu} \,
\lr{ 1 - \gamma^5}}

\threevertex[]{100}{300}{{vector}{$$}{
$\hat{A}^{\mu}_{(n)}$}}{{fermionout}{}{$f$}}{{fermionin}{}{$f$}}
{\quad \quad = \, i \, e_{(n)} \, Q_f \, \gamma^{\mu}}

\end{picture}
\end{center}
\mycaption{\label{FR} Feynman rules for couplings of the KK gauge
bosons to fermions in the minimal 5-dimensional extensions of the SM.}
\end{figure}

In Fig.~\ref{FR} we display the Feynman rules for the couplings of the
KK gauge bosons   to fermions   that   pertain to  the above   minimal
5-dimensional extensions of the SM.


\setcounter{equation}{0}
\section{Input Parameters, Observables and SM Predictions}
\label{observables}

In this Appendix, we list the numerical values of the input parameters
and electroweak observables,  along with  their SM predictions.  These
numerical values were used in Section~5 to constrain the parameters of
the 5-dimensional models.

As input parameters for  our theoretical predictions,  we use the most
accurately determined ones, namely the  Fermi constant $G_F$  measured
in muon decay, the fine  structure constant $\alpha$ determined by the
quantum Hall effect and the $Z$-boson mass $M_Z$ \cite{PDG}:
\begin{equation}
\label{inputparameters}
\begin{split}
G_F \ & = \ 1.16637(1) \, \times \, 10^{-5} \ \mr{GeV}^{-2} \, , \\
\alpha \ & = \ 1 / 137.0359895(61) \, , \\
M_Z \ & = \ 91.1872(21) \ \mr{GeV} \, ,
\end{split}
\end{equation}
where the numbers in parentheses indicate the $1\sigma$ uncertainties.

Given the above input parameters, predictions can be made for a number
of high-precision observables within the SM framework.  The results of
these   predictions    may  be  found     in~\cite{PDG}, together with
experimental values of the observables.  For reader's convenience, the
actual  values taken into account  in our analysis  are also listed in
Table~\ref{observablevaluesandsmpredictions}. The theoretical   values
in this table are obtained by assuming a light SM Higgs boson.

As was already discussed in Section~5,  we introduce an effective weak
mixing angle $\hat{\theta}_w$ by enforcing the tree-level SM relation
\begin{equation}
G_F \ = \ \frac{\pi \alpha}{\sqrt{2} \sin^2 \hat{\theta}_w \, \cos^2
\hat{\theta}_w \, M_Z^2} \ .
\end{equation}
If renormalization-group  running of   the  parameters   is  included,
e.g.~$\alpha (M_Z) \, = \, 1 / 128.92(3)$, we find
\begin{equation}
\sin^2 \hat{\theta}_w \ =\ 0.23105(8) \, ,
\end{equation}
which is the value used for the $Z$-pole observables in Section~5.

\begin{table}[fp]
\begin{center}
\small
\begin{tabular}{c|c|c} 
\hline\hline Observable & Exp. Value ($\mathcal{O}^{\mr{EXP}}$) & SM
Prediction ($\mathcal{O}^{\mr{SM}}$)\\ 
\hline \hline 
$M_W$ & 80.448(62) GeV & 80.378(20) GeV \\ \hline 
$\Gamma_Z (\mr{had})$ & 1.7439(20) GeV & 1.7422(15) GeV \\ \hline 
$\Gamma_Z (l^+ l^-)$ & 83.96(9) MeV & 84.00(3) MeV \\ \hline 
$\Gamma_Z (\nu \ov{\nu})$ & 498.8(15) MeV & 501.65(15) MeV \\ 
\hline $Q_W (\mr{Cs})$ & - 72.06(44)
             & -73.09(03) \\ \hline $R_e$ & 20.803(49) & 20.740(18) \\ \hline
$R_{\mu}$ & 20.786(33) & 20.741(18) \\ \hline 
$R_{\tau}$ & 20.764(45) & 20.786(18) \\ \hline 
$R_b$ & 0.21642(73) & 0.2158(2) \\ \hline 
$R_c$ & 0.1674(38) & 0.1723(1) \\ \hline 
$A_e$ & 0.15108(218) & 0.1475(13)\\ \hline 
$A_{\mu}$ & 0.137(16) & 0.1475(13) \\ \hline 
$A_{\tau}$ & 0.1425(44) & 0.1475(13) \\ \hline 
$A_b$ & 0.911(25) & 0.9348(1) \\ \hline 
$A_c$ & 0.630(26) & 0.6679(6) \\ \hline 
$A_s$ & 0.85(9) & 0.9357(1) \\ \hline 
$A_{\mr{FB}}^{(0,e)}$ & 0.0145(24) & 0.0163(3) \\ \hline 
$A_{\mr{FB}}^{(0,\mu)}$ & 0.0167(13) & 0.0163(3) \\ \hline
$A_{\mr{FB}}^{(0,\tau)}$ & 0.0188(17) & 0.0163(3) \\ \hline
$A_{\mr{FB}}^{(0,b)}$ & 0.0988(20) & 0.1034(9) \\ \hline
$A_{\mr{FB}}^{(0,c)}$ & 0.0692(37) & 0.0739(7) \\ \hline
$A_{\mr{FB}}^{(0,s)}$ & 0.0976(114) & 0.1035(9) \\ 
\hline\hline
\end{tabular}
\normalsize
\end{center}
\mycaption{\label{observablevaluesandsmpredictions} Precision
measurements and the corresponding SM predictions for all observables
considered in our analysis~\cite{PDG} (notation as in \cite{PDG}).}
\end{table}

\end{appendix}

\end{document}